\documentclass[useAMS,usenatbib,twocolumn]{mnras}
\usepackage{natbib}
\usepackage{graphicx}
\usepackage{xcolor}
\usepackage{times}
\usepackage{rotate}
\usepackage{color}
\usepackage{hyperref}
\usepackage{amsmath,amssymb,bm}
\usepackage{mathtools}
\usepackage{graphicx}
\usepackage{subfig}
\usepackage{lineno}
\usepackage{comment}
\usepackage{hyperref}

\newcommand{\DN}{$\mathrm{DN}4000$}
\newcommand{\wgg}{$w_\mathrm{gg}$}
\newcommand{\wgp}{$w_{\mathrm{g}+}$}

\newcommand{\Mpch}{\ensuremath{\mathrm{Mpc}~h^{-1}}}

\newcommand{\AIA}{ $A_\mathrm{IA}$ }

\begin{document}
  

\title[Intrinsic alignments with galaxy properties]{Intrinsic alignment demographics for next-generation lensing:\\Revealing galaxy property trends with DESI Y1 direct measurements} 

\author[Siegel, McCullough, Amon et al.]{
\parbox{\textwidth}{
\Large
J.~Siegel,$^{1}$\thanks{E-mail: siegeljc@princeton.edu}
J.~McCullough,$^{1}$
A.~Amon,$^{1,2}$
C.~Lamman,$^{3}$
N.~Jeffrey,$^{4}$
B.~Joachimi,$^{4}$
H.~Hoekstra,$^{5}$
S.~Heydenreich,$^{6}$
A.~J.~Ross,$^{7,8,9}$
J.~Aguilar,$^{10}$
S.~Ahlen,$^{11}$
D.~Bianchi,$^{12,13}$
C.~Blake,$^{14}$
D.~Brooks,$^{4}$
F.~J.~Castander,$^{15,16}$
T.~Claybaugh,$^{10}$
A.~de la Macorra,$^{17}$
J.~DeRose,$^{18}$
P.~Doel,$^{4}$
N.~Emas,$^{14}$
S.~Ferraro,$^{10,19}$
A.~Font-Ribera,$^{20}$
J.~E.~Forero-Romero,$^{21,22}$
E.~Gaztañaga,$^{15,23,16}$
S.~Gontcho A Gontcho,$^{10,24}$
G.~Gutierrez,$^{25}$
K.~Honscheid,$^{7,26,9}$
M.~Ishak,$^{27}$
S.~Joudaki,$^{28}$
R.~Kehoe,$^{29}$
D.~Kirkby,$^{30}$
T.~Kisner,$^{10}$
A.~Krolewski,$^{31}$
O.~Lahav,$^{4}$
A.~Lambert,$^{10}$
M.~Landriau,$^{10}$
L.~Le~Guillou,$^{32}$
M.~E.~Levi,$^{10}$
M.~Manera,$^{33,20}$
A.~Meisner,$^{34}$
R.~Miquel,$^{35,20}$
J.~Moustakas,$^{36}$
S.~Nadathur,$^{23}$
J.~ A.~Newman,$^{37}$
G.~Niz,$^{38,39}$
N.~Palanque-Delabrouille,$^{40,10}$
W.~J.~Percival,$^{41,31,42}$
A.~Porredon,$^{28}$
F.~Prada,$^{43}$
I.~P\'erez-R\`afols,$^{44}$
G.~Rossi,$^{45}$
E.~Sanchez,$^{28}$
C.~Saulder,$^{46}$
D.~Schlegel,$^{10}$
M.~Schubnell,$^{47,48}$
A.~Semenaite,$^{14}$
J.~Silber,$^{10}$
D.~Sprayberry,$^{34}$
Z.~Sun,$^{49}$
G.~Tarl\'{e},$^{48}$
B.~A.~Weaver,$^{34}$
R.~Zhou,$^{10}$
and H.~Zou$^{50}$
\begin{center} \textit{(DESI Collaboration)} \\\small{\textit{Affiliations are listed at the end of the paper}}\end{center}
}
}

\maketitle
\label{firstpage}

 
\begin{abstract} 
\noindent
We present direct measurements of the intrinsic alignments (IA) of over $2$~million spectroscopic galaxies using DESI Data Release 1 and imaging from four lensing surveys: DES, HSC, KiDS, and SDSS.
In this uniquely data-rich regime, we take initial steps towards a more tailored IA modelling approach by building a library of IA measurements across colour, luminosity, stellar mass, and redshift. We map the dependence between galaxy type---in terms of rest-frame colour, strength of the $4000~\mathrm{\AA}$ break, and specific star formation rate---and IA amplitude; the bluest galaxies have an alignment consistent with zero, across low ($0.05<z<0.5$) and high ($0.8<z<1.55$) redshifts. 
In order to construct cosmic shear samples that are minimally impacted by IA but maintain maximum sample size and statistical power, we map the dependence of alignment with colour purity.
Red, quenched galaxies are strongly aligned and the amplitude of the signal increases with luminosity, which is tightly correlated with stellar mass in our catalogues. For DESI galaxies between $0<z<1.5$,  trends in luminosity and colour alone are sufficient to explain the alignments we measure---with no need for an explicit redshift dependence.
In a companion paper (Jeffrey et al., in prep), we perform detailed modelling of the IA signals with significant detections, including model comparison. 
Finally, to direct efforts for future IA measurements, we juxtapose the colour-magnitude-redshift coverage of existing IA measurements against modern and future lensing surveys.
\end{abstract}


\section{Introduction}

As light travels towards us from distant galaxies, it is deflected by the intervening large scale structure of the Universe. 
This weak gravitational lensing is imprinted in the observed shapes of galaxies, and their coherent distortion, cosmic shear, is a powerful probe of the standard
cosmological model $\Lambda$CDM.
With increased survey size and improved methodology, the latest cosmic shear studies yield high-precision measurements of the cosmological parameters \citep{amon_2022,Secco2022,Dalal2023,Li2023twopoint,Wright2025}. 
Accounting for the correlations between galaxy shapes that have not been introduced by lensing (intrinsic alignment) is a limiting systematic to cosmic shear cosmology \citep[e.g.,][]{amon_2022}, alongside baryonic feedback \citep[e.g.,][]{Semboloni2011, Bigwood2024}.

As galaxies form and grow within halos, tidal effects are expected to influence their shapes and orientations \citep{Joachimi2015, Troxel2015, Lamman_2024}, by stretching \citep{Catelan2001} and/or torquing \citep{Schafer2009}.
The observed shapes of galaxies can therefore depend on their local environments.
This intrinsic alignment, IA, contaminates cosmic shear studies in two ways: i) physically close galaxies can have correlated shapes and ii) the shapes of galaxies in a foreground cluster are correlated with lensed background galaxies.
Cosmic shear studies commonly employ the Non-Linear linear-Alignment model \citep[NLA,][]{Catelan2001,Hirata2004,Bridle2007} to account for IA. NLA assumes the intrinsic ellipticities of galaxies are linearly related to the density field; the model also includes a non-linear correction to the linear matter power spectrum \citep{Bridle2007}, as well as luminosity, mass, and redshift dependence \citep{Joachimi2011}. However, this model is only valid on scales typically larger than those analysed in cosmic shear studies ($\gtrsim6$~Mpc~h$^{-1}$). The Tidal Alignment and Tidal Torquing model \citep[TATT,][]{Blazek2019} adds higher order terms, including tidal torquing, which extend the model validity to smaller scales ($\gtrsim2$~Mpc~h$^{-1}$), but the added flexibility comes at the cost of reduced cosmological precision. 
Additional models have been developed based on the halo model \citep{Fortuna2021} and effective field theory \citep{Vlah2020,Bakx2023,Chen2024,Maion2024};
these incur additional model parameters that exacerbate the loss of cosmological precision. 
In cosmic shear studies, models are typically applied without consideration of the properties of the lensing sample.
This can result in unnecessary degradation of cosmological precision if a highly flexible model is applied to a galaxy population with weak IA \citep{McCullough2024}.
To accurately and precisely account for IA in cosmic shear, we therefore require direct measurements to i) build informative priors on alignment strength as a function of galaxy type and redshift and ii) identify galaxy populations with minimal IA.

IA is directly measured by correlating galaxy shapes with their local environments as a function of separation, galaxy type, and redshift. 
Direct measurements require precise redshifts and shape measurements across wide areas of sky to correlate shapes and positions on scales between $0.1$ to $>100~{\Mpch}$.
These requirements have
limited prior studies to predominately bright, red galaxies at $z\lesssim1$.
With these limited samples, measurements have revealed strong alignments in ellipticals, with increased alignment in brighter galaxies \citep[e.g.,][]{Mandelbaum2006, Hirata2007,Joachimi2011,Singh2015,Johnston_2019,Fortuna2021,HervasPeters2024, Georgiou2025, CristinaFortuna2025,paus_ia, Xu_2023}.
The alignment amplitude of blue galaxies is less clear.
Direct measurements have yet to detect IA in blue galaxy samples \citep[e.g.,][]{Mandelbaum2011, Johnston_2019, Samuroff_2023, HervasPeters2024, Georgiou2025, paus_ia}; however, fewer measurements are available, and prior studies report wide constraints on the IA amplitude \citep[e.g., $A_1=0.07^{+0.32}_{-0.42}$ for the eBOSS ELGs with DES~Y3 imaging,][]{Samuroff_2023}.
Direct IA measurements across the entire colour, redshift, and luminosity space sampled by modern weak lensing surveys is critical to building an accurate census and informing next generation modelling techniques.   

In this paper, we present direct IA measurements with spectroscopic redshifts from the first year of the DESI Main Survey (DESI~DR1) and shape measurements from four Stage III lensing surveys: DES, KiDS, HSC, and SDSS.
With over $4$~million galaxies across $>9,000$~deg$^2$, DESI~DR1 is well poised for IA measurements.
Prior measurements predominantly relied on SDSS and GAMA spectroscopy, resulting in comparatively smaller samples with overlapping redshift and shape measurements, or photometric redshifts, which introduce additional uncertainties.
Across the four lensing surveys we consider, there are over $2$~million galaxies with DESI~DR1 spectroscopy, approximately a factor of 20 more than previous studies. 
In addition to larger sample sizes, DESI spectroscopic samples also cover a colour, luminosity, and redshift space more relevant to modern weak lensing surveys. 
Leveraging these advantages, we present direct IA measurements for a wide range of galaxy populations between the local Universe and $z\sim1.5$.
These measurements will inform the modelling of intrinsic alignments in current and next generation cosmic shear studies.
We additionally report optimal selections of weakly aligned galaxies.

We present the spectroscopic and imaging data in Section~\ref{sec:data}.
Our measurement methods and results are reported in Sections~\ref{sec:2pt}.
We discuss the optimal selection of weak IA galaxies in Section~\ref{sec:characterizing_for_shear} and investigate the physical drivers of alignment in Section~\ref{sec:physical_drivers}. 
We review future prospects in Section~\ref{sec:landscape} and conclude in Section~\ref{sec:conclusions}.
For the sake of comoving distance calculations, we adopt the cosmological parameters: $\Omega_{\rm m} = 0.3$, and $\Omega_\Lambda = 0.7$. 
We report comoving distance in ${\Mpch}$.

\begin{figure*}
\includegraphics[width=\textwidth]{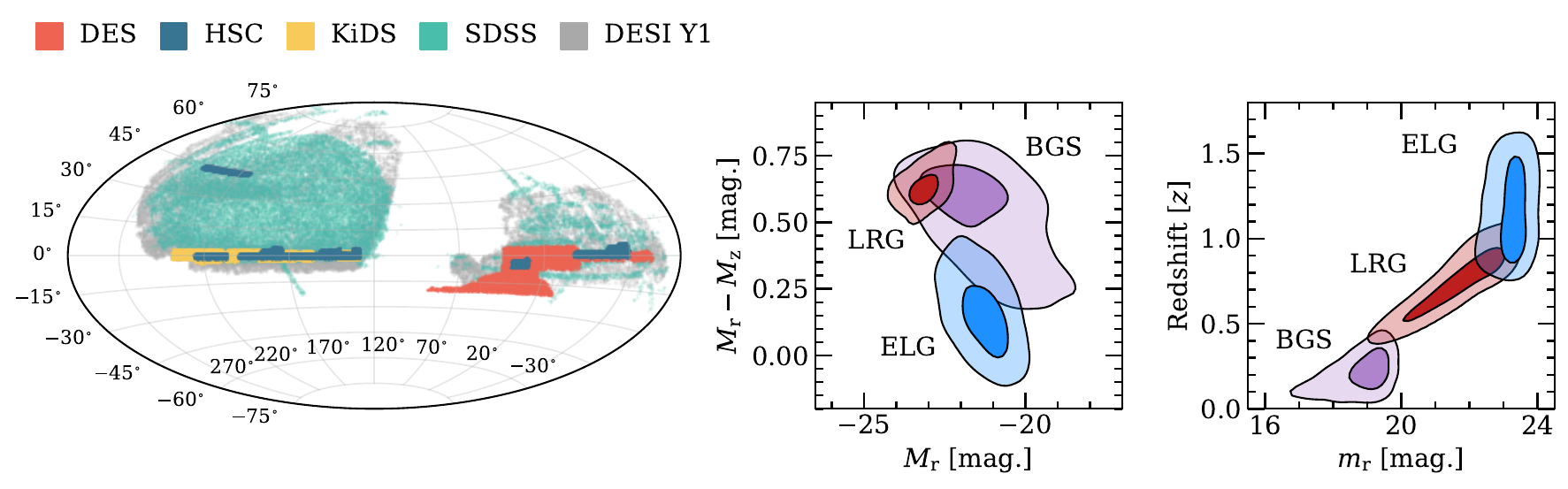}
\caption{The leftmost panel presents the footprint of the DESI DR1 data (the BGS, LRG, and ELG samples are shown in grey) alongside the four imaging surveys: DES (red), KiDS (yellow), HSC (blue), and SDSS (green);
the DES and KiDS footprints are truncated to declinations overlapping with DESI.
The overlap area and properties of the matched data between each DESI sample and each imaging survey are summarized in Table~\ref{tab:data_cross_matching}. 
The center and rightmost panels show the properties of the DESI DR1 BGS (purple), LRG (red), and ELG (blue)  galaxy samples;
the galaxies' rest-frame magnitudes, rest-frame colours, apparent magnitudes, and redshifts are presented. 
The contours indicate the $39$ and $86$th percentiles. 
} 
\label{fig:onsky}
\end{figure*}

\section{Data} \label{sec:data}

We measure IA using spectroscopic redshifts from the DESI DR1 data assembly and shape measurements from four imaging surveys: DES, KiDS, HSC, and SDSS; 
the shear catalogues are cross-matched to the DESI redshifts. 
The footprints of each survey are presented in Figure~\ref{fig:onsky}.
We outline the imaging and spectroscopic surveys in Sections~\ref{sec:data_imaging} and ~\ref{sec:data_desi}, respectively. 
In Section~\ref{sec:data_samples}, we present our redshift cross-matched shape samples, including sub-populations designed to study the colour, luminosity, and redshift dependence of the intrinsic alignment signal.

\begin{table*}
\begin{center}
\begin{tabular}{llc|cccccccccc}
\hline
Imaging & Sample & No. of Galaxies & $f_{\rm match}$ & Area & $\bar{n}$ & $\sigma_{\rm e}$ & $\mathcal{R}$& $\bar{m}$ & Redshift & Colour & Luminosity  \\
 &  &  &  & [deg.$^2$] & [deg.$^{-2}$] &  & & & $\langle z \rangle$  & $\langle M_r-M_z \rangle$  & $\langle M_r \rangle$  \\
\hline
DES & BGS & 202,244 & $0.745$ & $701$ & $288.4$ & $0.16$ & $0.755$ & $-0.006$ & $0.24$ & $0.53$ & $-21.43$\\
 & LRG & 144,187 & $0.82$ & $802$ & $179.6$ & $0.14$ & $0.675$ & $-0.024$ & $0.71$ & $0.63$ & $-23.18$\\
 & ELG & 65,316 & $0.362$ & $742$ & $87.9$ & $0.19$ & $0.632$ & $-0.037$ & $1.12$ & $0.17$ & $-21.71$\\
\hline
KiDS & BGS & 229,510 & $0.855$ & $447$ & $513.2$ & $0.26$ & --- & $0.287$ & $0.22$ & $0.54$ & $-21.39$\\
 & LRG & 151,467 & $0.774$ & $444$ & $340.4$ & $0.22$ & --- & $0.107$ & $0.72$ & $0.64$ & $-23.14$\\
 & ELG & 136,706 & $0.449$ & $447$ & $305.5$ & $0.25$ & --- & $-0.172$ & $1.13$ & $0.17$ & $-21.6$\\
\hline
HSC & BGS$^*$ & 31,102 & $0.123$ & $436$ & $71.3$ & $0.26$ & $0.847$ & $-0.072$ & $0.36$ & $0.59$ & $-22.66$\\
 & LRG & 131,533 & $0.742$ & $432$ & $304.1$ & $0.29$ & $0.865$ & $-0.068$ & $0.76$ & $0.63$ & $-23.14$\\
 & ELG & 162,267 & $0.634$ & $433$ & $374.2$ & $0.38$ & $0.853$ & $-0.055$ & $1.14$ & $0.17$ & $-21.38$\\
\hline
SDSS & BGS & 1,608,722 & $0.827$ & $4797$ & $335.3$ & $0.36$ & $0.869$ & --- & $0.22$ & $0.54$ & $-21.47$\\
 & LRG$^*$ & 387,373 & $0.368$ & $4365$ & $88.7$ & $0.38$ & $0.853$ & --- & $0.61$ & $0.65$ & $-23.16$\\
 & ELG$^*$ & 2,005 & $0.002$ & $4060$ & $0.5$ & $0.51$ & $0.739$ & --- & $0.98$ & $0.2$ & $-22.26$\\
\hline
\end{tabular}
\caption{
The properties of the shape tracer samples used in this work.
We report: 
the total number of DESI to imaging matched galaxies;
$f_\mathrm{match}$, the number of galaxies in the matched sample relative to the parent DESI sample (within a $10$~deg.$^2$ area where the imaging and DESI footprints overlap);
the area of the footprint in square degrees;
$\bar{n}$, the mean surface density of galaxies;
$\sigma_{\epsilon}$, the ellipticity dispersion  \citep{Heymans2012};
$\mathcal{R}$, the shear responsivity for each sample; the multiplicative shear calibration factor, $m$;
and the weighted mean redshift $\langle z \rangle$, absolute colour $\langle M_r-M_z \rangle$, and absolute magnitude $\langle M_r \rangle$. 
$^*$We omit the BGS--HSC, LRG--SDSS, and ELG--SDSS samples from our measurements due to the low match fraction between the density and shape catalogues.
}
\label{tab:data_cross_matching}
\end{center}
\end{table*}

\subsection{Imaging}
\label{sec:data_imaging}

The on-sky projected shape of a galaxy is described by its complex ellipticity $\epsilon$ (the amplitude and orientation of the major and minor axes).
Because IA measurements are informative for cosmic shear, we adopt the shape conventions and calibrations of weak lensing surveys: observed shape measurements $\epsilon^\mathrm{obs}$ are related to the shear estimate $\gamma$ as \citep{Kaiser1995,Bernstein2002}
\begin{equation}
    \gamma = \frac{1}{1+m} \left ( \frac{\epsilon^\mathrm{obs}}{\mathcal{R}_\mathrm{eff} } + c \right ),
\end{equation}
where $m$ and $c$ are the multiplicative and additive biases, respectively.
Image noise, modelling bias, detection bias, and PSF size misestimation are all potential sources of multiplicative biases $m$, while additive biases $c$ can arise from misestimated PSF asymmetries, or an incomplete correction of the PSF anisotropy. 
$\mathcal{R}$ is the effective responsivity (the response of the ellipticity to a small distortion).
When shapes are measured via
polarization  $\mathcal{R}_\mathrm{eff} = 2 \mathcal{R}$ (e.g., HSC and SDSS), otherwise $\mathcal{R}_\mathrm{eff} = \mathcal{R}$ \citep{Mandelbaum2006}.
For all surveys, we apply the ellipticity calibrations at the level of the two-point correlation (see Table~\ref{tab:data_cross_matching} for each sample's calibration coefficients).

\subsubsection{DES}
\label{sec:DES_imaging}

Over a six year program, the Dark Energy Survey (DES) imaged $5000~\mathrm{deg}^2$ of the Southern sky ($grizY$) with the Blanco $4$~meter telescope at Cerro Telolo Inter-American Observatory \citep[]{DES2005}.
We consider the DES Year~3 (Y3) shear catalogue,\footnote{\url{https://des.ncsa.illinois.edu/releases/y3a2}} which includes $100$~million galaxies over nearly $4000~\mathrm{deg}^2$ to $i<23$~mag \citep{SevillaNoarbe2021,Gatti_2021}.   

DES Y3 ellipticities are measured via \textsc{Metacalibration}, in which the shape measurement biases are inferred by artificially shearing the observed images \citep{Huff2017, Sheldon2017, Gatti_2021}.
The shear response matrix $\bm{\mathcal{R}}$ is calculated per galaxy by considering the ellipticity measurements of each artificially sheared image:
\begin{equation}
    \bm{\mathcal{R}}_{i,j} = \frac{ \epsilon_i^{ s_{j+} } - \epsilon_i^{ s_{j-} } }{ \Delta \gamma_j },
\end{equation}
where $\epsilon_i^{ s_{j+/-} }$  is the measured ellipticity of the $i$th component on the image positively (negatively) sheared by $ \Delta \gamma_j$ in the $j$th component;
the full response matrix has been shown to be well approximated by $\mathcal{R} = ( \bm{\mathcal{R}}_{11} + \bm{\mathcal{R}}_{22} )/2$ \citep[see Appendix A of][]{Gatti_2021}.
For a given galaxy sample, we calculate the average responsivity as
\begin{equation}
    \mathcal{R} = \frac{\sum_{i=1}^N  w_i (\bm{\mathcal{R}}_{i,11}+\bm{\mathcal{R}}_{i,22})/2 }{ \sum_{i=1}^N  w_i },
\end{equation}
where $w_i$ is the inverse-variance weight of the $i$th galaxy and $N$ is the number of galaxies.

Biases not handled by \textsc{Metacalibration} (e.g., shear-dependent detection and blending) are calibrated by image simulations \citep{maccrann_2022}.
We adopt the redshift dependent multiplicative biases from the DES Y3 cosmic shear analysis \citep[$m \sim 2\%$,][]{amon_2022}.

\subsubsection{KiDS}
\label{sec:kids_imaging}

The Kilo Degree Survey (KiDS; \citealp{dejong2013}) was carried out at Paranal from 2011 to 2019 using OmegaCAM on the VLT Survey Telescope. The final survey covers about 1350 deg$^2$, but here we use the penultimate data release (DR4\footnote{\url{https://kids.strw.leidenuniv.nl/DR4/index.php}}; \citealp{kuijken2019}) which covers about $1000\,{\rm deg}^2$ before masking (only the equatorial field overlaps with DESI; see Figure~\ref{fig:onsky}). The galaxy ellipticity measurements in the public KiDS catalogues are based on \textit{lens}fit \citep{Miller2013, FenechConti2017}, which  
includes a bright magnitude cut of $m_r=20$. 
This effectively eliminates any overlap with the DESI BGS sample (see Section~\ref{sec:bgs}). 
To remedy this, we remeasure galaxy ellipticities from the $r$-band imaging (mean seeing of $0.7^{\prime\prime}$) using a bespoke pipeline similar to \citet{johnston2021}; the simulated data were adjusted to match the KiDS data in terms of passband, image quality, and depth.

Galaxy ellipticities were measured by weighted moments of each galaxy's observed surface brightness distribution. The quadrupole moments were corrected for PSF effects via the KSB method \citep{Kaiser1995,luppino1997,hoekstra1998}. 
The weight function used to compute the moments determines which scales in the galaxy image the shear estimate is most sensitive to; the alignment amplitude can depend on this choice \citep{Georgiou2019, Georgiou2025}. 
Cosmic shear studies generally use weight functions that match the observed surface brightness profiles. Therefore, we adopted an axisymmetric Gaussian weight function with a dispersion 
$r_{\rm g}=\sqrt{2\ln 2}r_{\rm h}$, where $r_{\rm h}$ is the half-light radius measured by \texttt{SourceExtractor} \citep{Bertin1996}. 

The resulting ellipticity estimates are biased by noise and imperfections in the corrections. However, these can be accurately accounted for by calibrating with image simulations \citep{Hoekstra2015}. We created simulated KiDS images following \citet{Hoekstra2015, Hoekstra2017}. The input galaxies 
are modelled as single S{\'e}rsic profiles, with properties extracted from the GEMS survey \citep{rix2004}. 
Noise levels in the images were set to match KiDS $r$-band data. 
Fifty PSFs were drawn randomly from the survey; these were modelled as Moffat profiles, similar to \cite{Li2023}. 
Using these realistic PSFs we also quantify the residual additive bias arising from imperfections in the correction for PSF anisotropy. 
Following \cite{Hoekstra2021}, we used the simulations to determine a  `boost' factor for the smear polarisability.
This improvement was not considered in \cite{johnston2021} and effectively removes the residual additive bias.
The multiplicative bias is assumed to affect the two shear components equally and was calculated as a function of the angular size of the galaxy, its signal-to-noise ratio, and the PSF size. 
The resulting multiplicative biases are larger than those seen for the other weak lensing surveys (see Table~\ref{tab:data_cross_matching}), because of the choice of shape measurement method. In particular, the KSB algorithm does not account for the pixel noise in the image, which biases the shapes low \citep{kacprzak2012,viola2014, Hoekstra2015}, resulting in negative values for $m$ for faint galaxies. For large, bright galaxies the adopted weight function results in a positive value for $m$.
For the bespoke calibration, responsivity is encapsulated in the multiplicative bias; we therefore fix $\mathcal{R_\mathrm{eff}}=1$. 

\subsubsection{HSC}
\label{sec:hsc_imaging}

The Hyper Suprime-Cam (HSC) Subaru Strategic Program is conducting a 300-night multi-band imaging survey with the $8.2$~meter Subaru telescope \citep{Aihara2018}.
We use the three year galaxy shear catalogue,\footnote{\url{https://hsc-release.mtk.nao.ac.jp/doc/}} which covers $416$~deg$^2$ to $i < 24.5$~mag and $0.3 \lesssim z \lesssim 1.5$ \citep{Li2022}.
Shapes are measured with the \texttt{GalSim} \citep{Rowe2015} re-Gaussianization PSF correction method \citep{Hirata2003}.

Following the HSC Y3 cosmic shear analysis \citep{Dalal2023}, we approximate the shear responsivity as
\begin{equation}
    \mathcal{R} = 1 - \frac{ \sum_{i=1}^{N} w_i e_\mathrm{i,rms}^2 }{ \sum_{i=1}^{N} w_i },
\end{equation}
where $e_{i,\mathrm{rms}}$ is the RMS of the intrinsic ellipticity for the $i$th galaxy, $w_i$ is the per-galaxy inverse variance weight (composed of the photon noise shape uncertainty and $e_\mathrm{rms}$), and $N$ is the number of galaxies \citep{Hirata2003}.
Multiplicative biases are estimated for each galaxy using Hubble Space Telescope observations overlapping the HSC footprint \citep{Mandelbaum2018, Li2022}.
The weighted average multiplicative bias is calculated for each measurement sample.

\subsubsection{SDSS}
\label{sec:sdss_imaging}

The Sloan Digital Sky Survey \citep[SDSS,][]{York2000} imaged over $10,000$~deg$^2$ in five optical bands ($ugriz$).
We consider the flux-limited ($r < 22$ and $i < 21.6$) shape catalogue of \cite{Reyes2012}, derived from the seventh SDSS data release \citep{Abazajian2009};
the shape catalogue includes $43$~million galaxies over $9493$~deg$^2$. 
To facilitate survey-to-survey comparison and co-adding, we mask the regions of the SDSS footprint that overlap with DES and KiDS.

Ellipticities are measured via adaptive moments \citep[see Appendix A of][]{Reyes2012}.
Following \cite{Reyes2012} and \cite{Mandelbaum2013}, we approximate the shear responsivity as 
\begin{align}
    \mathcal{R} &\approx 1 - e_\mathrm{rms}^2,\\
    e_\mathrm{rms}^2 &= \frac{1}{2N} \sum_{i=1}^{N} ( e_{+,i}^2 +  e_{\times,i}^2 ),
\end{align}
where $w_i$ is the per-galaxy inverse variance weight and $N$ is the number of galaxies in the sample. 
No multiplicative bias correction is readily available.

\subsection{DESI}
\label{sec:data_desi}

The Dark Energy Spectroscopic Instrument (DESI) is a ground-based multi-fiber spectrograph installed at the $4$~meter Mayall telescope \citep{DESIexperiment2016,Schlafly2023,DESICollaboration2024SV,DESICollaboration2024EDR}.
Fed by $5,020$ reconfigurable fibers, DESI covers the optical to near-infrared---$380$ to $960$~nm---in a single exposure \citep{DESICollaborationInst2016,DesiCollaboration2022,Silber2023,Miller2024,Poppett2024}.
To constrain the Universe's expansion history and the
growth rate of large scale structure \citep{Levi2013}, DESI is undertaking a five year $14,200$~deg$^2$ survey of five extragalactic tracers: bright galaxies in the magnitude limited bright galaxy survey (BGS),
luminous red galaxies (LRG), 
emission line galaxies (ELG), 
quasars, and
Lyman-$\alpha$ forest quasars. 
In this work, we consider the BGS, LRG, and ELG DESI~DR1 large-scale structure catalogues \citep{desi_2pt_clustering, Ross2025,DESIDR12025}.\footnote{\url{https://data.desi.lbl.gov/public/dr1/survey/catalogs/dr1/LSS/iron/LSScats}}

DESI spectroscopic redshifts are derived from the one-dimensional spectra \citep{Guy2023} with \texttt{Redrock}\footnote{\url{https://redrock.readthedocs.io}} (Bailey et al., in prep.).
\texttt{Redrock} determines the best-fit redshift by minimizing the $\chi^2$ between the observed spectrum and a combination of spectral templates.
Restframe luminosities and colours are inferred from the one-dimensional DESI spectra and DESI Legacy Imaging Surveys \citep{Zou2017,Dey2019} with \texttt{FastSpecFit}\footnote{\url{https://github.com/desihub/fastspecfit}} \citep{fastspecfit} and \texttt{CIGALE} \citep{Siudek2024}.

To account for survey systematics, we also consider the DESI~DR1 large scale structure random catalogues.
The randoms sample the area on the sky where DESI~Y1 data could have been observed and match the radial distribution of the observations by sampling redshifts directly from the data; see \cite{Ross2025} for more details.

\begin{table}
\begin{center}
\begin{tabular}{ll|ccc}
\hline
Sample & Imaging & Reduced $\chi^2$ & $p(>\chi^2)$ &  Sig. ($\sigma$)\\
\hline
BGS & DES & $2.8$ & $2.5\times10^{-2}$ & $2.2$\\
 & KiDS & $1.6$ & $1.6\times10^{-1}$ & $1.4$\\
 & SDSS & $11.9$ & $<10^{-6}$ & $6.1$\\
& Co-added & $20.4$ & $<10^{-6}$ & $>10$\\
\hline
LRG & DES & $1.9$ & $1.1\times10^{-1}$ & $1.6$\\
 & KiDS & $8.5$ & $<10^{-6}$ & $5.0$\\
 & HSC & $2.5$ & $4.0\times10^{-2}$ & $2.0$\\
& Co-added & $31.8$ & $<10^{-6}$ & $>10$\\
\hline
ELG & DES & $3.4$ & $8.5\times10^{-3}$ & $2.6$\\
 & KiDS & $0.8$ & $5.1\times10^{-1}$ & $0.7$\\
 & HSC & $1.8$ & $1.4\times10^{-1}$ & $1.5$\\
& Co-added & $1.1$ & $3.6\times10^{-1}$ & $0.9$\\
\hline
\end{tabular}
\caption{
The significance of the measured IA correlations between each DESI density tracer and shear catalogue, as well as the cross-survey co-added measurements; 
HSC is omitted from co-adding, due to its significant overlap with the other imaging surveys.
Reduced $\chi^2$ is calculated relative to the null hypothesis of zero signal between $6-65~{\Mpch}$ (corresponding to $4$ degrees of freedom).
For reference, the $\chi^2$ test p-value is also reported in terms of standard deviations for a one-sided Gaussian.
}
\label{tab:detection}
\end{center}
\end{table}

\subsection{Galaxy Samples}
\label{sec:data_samples}

Direct measurements of IA require a \textit{density} catalogue with precise redshifts and a \textit{shape} catalogue with precise shape and redshift information.
To construct our \textit{shape} samples, we cross-match the DESI redshift catalogues (BGS, LRG and ELG) with the DES, KiDS, HSC, and SDSS shear catalogues (with a matching tolerance of $1''$).
We characterize the mean colour, luminosity, and redshift of the shape catalogues in Table~\ref{tab:data_cross_matching}, with an extended comparison provided in Appendix~\ref{sec:crossmatching_appendix}.

\subsubsection{BGS}
\label{sec:bgs}

The complete bright galaxy survey (BGS) will consist of over $14$~million galaxies between $0 < z < 0.6$.
The design of the BGS is presented in \cite{Hahn2023BGS} and is briefly described below.
The survey consists of a $r < 19.5$~mag. bright
sample (BGS Bright) and a $19.5 < r < 20.175$~mag. fainter sample (BGS Faint); 
the faint sample includes a series of colour cuts to optimize redshift recovery. 
We consider the BGS Bright galaxies in the redshift range $0.05 < z < 0.5$;
our DESI DR1 sample consists of approximately $4,000,000$ galaxies with $\langle r \rangle=18.7$ [mag.], $\langle r-z \rangle=0.67$ [mag.], and $\langle z \rangle=0.21$.

The BGS shape catalogues are summarized in Table~\ref{tab:data_cross_matching}.
With the exception of HSC, the match fraction is high between DESI and the imaging surveys ($\gtrsim 80\%$) with little dependence on galaxy luminosity or restframe colour (Figure~\ref{fig:samples_2d}).
We omit the BGS--HSC sample from our measurements due to the low match fraction: $12\%$.
The BGS to HSC cross-matching is limited by the omission of low-redshift galaxies from the HSC shear catalogue.

To study the colour and luminosity dependence of the intrinsic alignment signal, we divide the BGS shape catalogues by rest-frame colour and luminosity (in terms of the absolute $r$-magnitude):
\begin{enumerate}

    \item Blue: $M_\mathrm{r} - M_\mathrm{z}< 0.5$ 
    \begin{enumerate}
        \item Faint:  $~~M_r > - 20.9$
        \item Bright: $M_r < - 20.9$
    \end{enumerate}
    \item Red:  $M_\mathrm{r} - M_\mathrm{z}> 0.5$
    \begin{enumerate}
        \item Faint:  $~~M_r > - 21.9$
        \item Bright: $M_r < - 21.9$
    \end{enumerate}
\end{enumerate}

\subsubsection{LRG}

The luminous red galaxy (LRG) survey will observe $8$~million galaxies between $0.4 < z < 1.1$.
The LRG survey is presented in \cite{Zhou2023LRG} and is briefly described below.
Through a series of optical ($grz$) and infrared (WISE W1) magnitude and colour cuts, the LRG survey targets bright galaxies with strong $4000~\mathrm{\AA}$ breaks.
We use the full LRG redshift range ($0.4 < z < 1.1$). Our DESI DR1 LRG sample consists of approximately $2,100,000$ galaxies with $\langle r \rangle=21.6$ [mag.], $\langle r-z \rangle=1.6$ [mag.], and $\langle z \rangle=0.74$.

The LRG shape catalogues are summarized in Table~\ref{tab:data_cross_matching}.
Other than SDSS, the match fractions are high ($\gtrsim 75\%$) with little dependence on galaxy luminosity or restframe colour (Figure~\ref{fig:samples_2d}).
We omit the LRG--SDSS sample from our measurements due to the low match fraction ($35\%$) and the matching's strong dependence on redshift.

We divide the LRG sample into two redshift bins and two luminosity bins for each redshift range:
\begin{enumerate}

    \item Low redshift: $0.4 < z < 0.75$
    \begin{enumerate}
        \item Faint:  $~~M_r > - 23.1$
        \item Bright: $M_r < - 23.1$
    \end{enumerate}
    \item High redshift: $0.75 < z < 1.1$
    \begin{enumerate}
        \item Faint:  $~~M_r > - 23.3$
        \item Bright: $M_r < - 23.3$
    \end{enumerate}
    
\end{enumerate}

\subsubsection{ELG}
The emission line galaxy (ELG) survey is targeting $30$~million star forming galaxies between $0.6 < z < 1.6$, presented in \cite{Raichoor2023ELG}. 
The samples were selected from optical ($grz$) photometry.
Both samples require $g > 20$ and $g_\mathrm{fib}<24.1$ (where $g_\mathrm{fib}$ is the fiber adjusted $g$-band magnitude), in addition to $g - r$ and $r - z$ cuts; to optimize redshift recovery, the colour cuts vary between the low and high redshift samples.
We consider ELG targets between $0.8 < z < 1.55$. 
Our DESI DR1 ELG sample consists of approximately $2,400,000$ galaxies with $\langle r \rangle=23.2$ [mag.], $\langle r-z \rangle=0.68$ [mag.], and $\langle z \rangle=1.2$.

The ELG shape catalogues are summarized in Table~\ref{tab:data_cross_matching}.
With the exception of SDSS, the match fractions are $\gtrsim 40\%$ with minor dependence on galaxy luminosity and restframe colour (Figure~\ref{fig:samples_2d}); there is a slight preference for brighter galaxies.
We omit the ELG--SDSS sample from our measurements due to the $0.2$\% match fraction;
the ELG to SDSS matching suffers from the shallower depth of SDSS imaging.

Analogous to the LRG sample, we divide the ELG galaxies into two redshift bins and two luminosity bins for each redshift range:
\begin{enumerate}

    \item Low redshift: $0.8 < z < 1.15$
    \begin{enumerate}
        \item Faint:  $~~M_r > - 21.3$
        \item Bright: $M_r < - 21.3$
    \end{enumerate}
    \item High redshift: $1.15 < z < 1.55$
    \begin{enumerate}
        \item Faint:  $~~M_r > - 21.9$
        \item Bright: $M_r < - 21.9$
    \end{enumerate}
    
\end{enumerate}

\section{Methodology} \label{sec:2pt}

\subsection{Two-point correlation function measurements}

We measure the intrinsic alignment of galaxies with the projected two-point cross-correlation function $w_{\rm g+}$.
Following \cite{Mandelbaum2006}, we first calculate the three-dimensional correlation function with the modified Landy--Szalay estimator \citep{Landy1993}:
\begin{align}
    \xi_{\rm g+}(r_{\rm p}, \Pi) &= \frac{ S_+ (D - R) }{ R R },\\
    S_+ D &= \sum_{i\neq j \mid r_{\rm p}, \Pi} \gamma_+(i\mid j).
\end{align}
$ S_+ D$ is the sum of the tangential shear component between all galaxy pairs in the shape and density samples with i) the comoving transverse component $x_\perp$ satisfying $\mid \log_{10}(r_{\rm p}) - \log_{10}(x_\perp) \mid <  \Delta \log(r_{\rm p})/2$, where $\Delta \log(r_{\rm p})$ is a constant log bin width, and ii) the comoving radial component $x_\parallel$ satisfying  $ \mid x_\parallel - \Pi \mid < \Delta \Pi /2$, where $\Delta \Pi$ is a constant linear bin width.
Analogously, $ S_+ R$ is the sum of the tangential shear component between the shape and random catalogues.
$RR$ is the number of pairs in the random catalogue within the $r_{\rm p}$ and $\Pi$ bins.
By the same process, we also measure $\xi_{g\times}(r_{\rm p}, \Pi)$, where the galaxy shears are rotated by $\pi / 4$; this is a valuable systematics test, because we expect the $\xi_{g\times}(r_{\rm p}, \Pi)$ to be consistent with zero by parity.

We measure galaxy-galaxy clustering with the \cite{Landy1993} estimator:
\begin{align}
    \xi_{\rm gg}(r_{\rm p}, \Pi) &= \frac{ (D-R) (D - R) }{ R R },
\end{align}
where $DR$ is the number of pairs between the density and random catalogues within the $r_{\rm p}$ and $\Pi$ bin. 

By integrating the three-dimensional correlation functions along the line of sight, we arrive at the projected correlation function:
\begin{equation}
    \label{eqn:twopoint}
    w_{\rm ab} (r_{\rm p}) = \int_{-\Pi_\mathrm{max}}^{\Pi_\mathrm{max}} \xi_{\rm ab}(r_{\rm p},\Pi) d \Pi,
\end{equation}
where $ab \in (g+, g\times, gg) $ and the integration is between $-\Pi_\mathrm{max}$ and $\Pi_\mathrm{max}$.

The $w_{\rm g+}$ and $w_{\rm g\times}$ two-point correlation functions are measured with \texttt{TreeCorr}\footnote{\url{http://rmjarvis.github.io/TreeCorr}} \citep{treecorr} and the publicly available \cite{Johnston_2019} pipeline.\footnote{\url{https://github.com/harrysjohnston/2ptPipeline}}
We use $12$ logarithmically spaced bins between $0.2-200~{\Mpch}$ for the comoving transverse component $x_\perp$ and $14$ linearly spaced bins between $-60$ and $60~{\Mpch}$ for the comoving line of sight component $x_\parallel$.
Measurement covariance matrices are estimated via a leave-one-out jackknife process, described in Appendix~\ref{sec:covariances_appendix}.

The {\wgg} correlation function (identical to the classic 2-point projected galaxy correlation function) is measured with the DESI large-scale structure (LSS) pipeline \citep{desi_2pt_clustering}, which builds on \texttt{pycorr}\footnote{\url{https://github.com/cosmodesi/pycorr}} \citep{Sinha2019,Sinha2020}.

We validated the modelling and measurement pipelines with \textsc{buzzard} mock catalogues \citep{derose_2019} adapted for use in DESI \citep{Lange_2024}, as well as against previous measurement pipelines in \cite{Johnston_2019, Samuroff_2023, Lamman_2024}.

We include per galaxy weights when calculating the correlation functions, to account for differences in signal-to-noise ratio and survey systematics. 
For the density tracers, we adopt the recommended large scale structure weights \citep[\texttt{WEIGHT},][]{Ross2025}.
The inclusion of angular upweighting and pairwise inverse probability weights, which account for the effects of fiber collisions (see Appendix C of \cite{desi_2pt_clustering} and \cite{Bianchi2025}), does not impact our results.
The shape tracers are weighted by the product of the DESI LSS weights and the shear catalogue weights.

For each DESI sample (BGS, LRG, ELG), we measure the two-point shape--density correlation function between the DESI density tracers and shape catalogues from four shear surveys (DES, HSC, KiDS, SDSS).
The IA measurements are consistent between the lensing surveys (Appendix~\ref{sec:survey_to_survey_appendix}). The galaxy-galaxy lensing signals from the same data were also found to be consistent \citep{heydenreich2025}.

\begin{figure*}
    \includegraphics[width=0.8\textwidth]{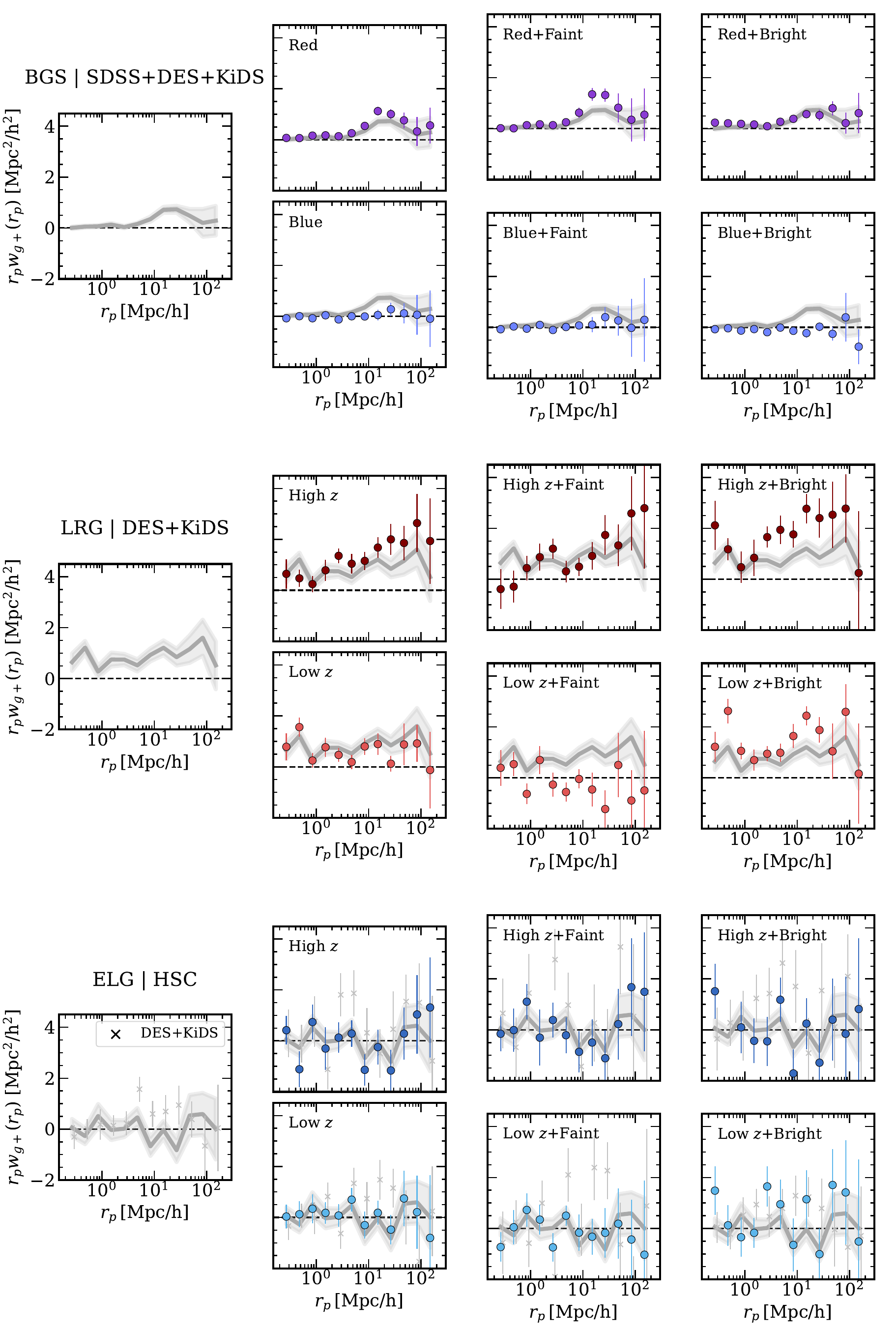}
    \caption{ 
    Measured DESI IA correlations: the alignment signals from the BGS, LRG, and ELG cross-matched shear catalogues (co-added across multiple lensing surveys) are presented in the left-most column; the choice of surveys for co-adding is discussed in Section~\ref{sec:modelling}.
    BGS (upper panel) is divided into blue ($M_r-M_z<0.5$) and red ($M_r-M_z>0.5$) subsamples in the second column, and further into faint and bright subsamples by luminosity in the right column.  
    The LRG sample (middle panel) is divided by redshift at $z=0.75$ (middle column) and further by luminosity (right column). 
    Similarly, the ELGs (bottom panel) are divided by redshift at $z=1.15$ (middle column) and further by luminosity (right column). 
    As a visual guide, the signal from the parent DESI sample (i.e., no colour, redshift, or luminosity cuts) is shown in grey alongside all of the sub-samples.
    } 
    \label{fig:subsample_summary}
\end{figure*}

\begin{figure*}
    \includegraphics[width=0.9\textwidth]{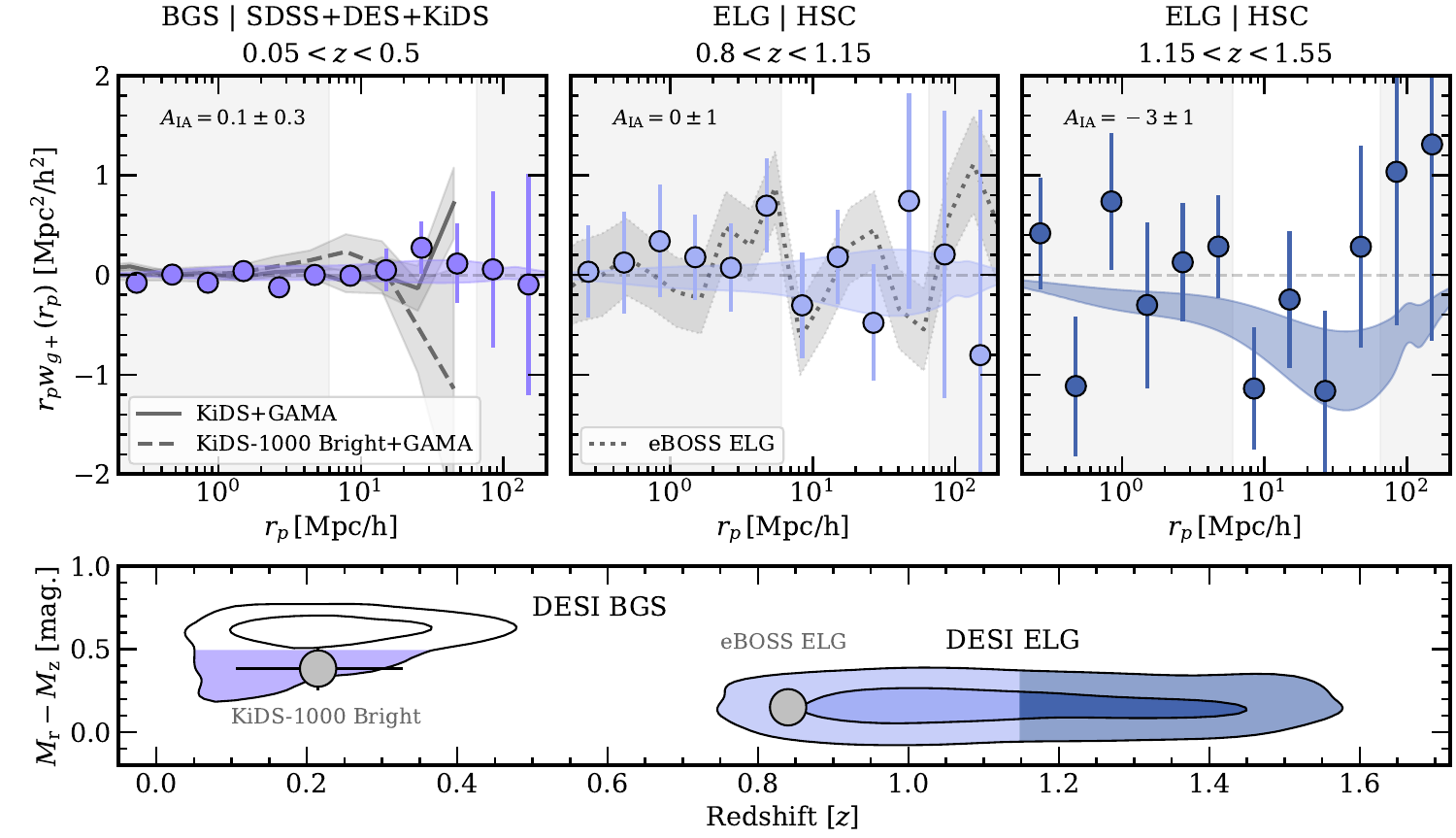}
    \caption{ 
    Blue star-forming galaxies are consistent with zero intrinsic alignments, independent of redshift.
    Top left: the IA signal for blue ($M_\mathrm{r} - M_\mathrm{z}<0.5$) BGS galaxies (co-added across the SDSS, DES, and KiDS shape catalogues).
    Top center and right: the IA signal for low ($z<1.15$) and high redshift ($z>1.15$) ELG galaxies from the HSC shape catalogue; the DES$+$KiDS ELG signal is also null (see Appendix~\ref{sec:survey_to_survey_appendix}).
    The $16$th and $84$th percentiles of an NLA fit to the measurements are shown as blue shaded bands and the inferred $A_\mathrm{IA}$ are reported in the upper left of each panel. The measurements are fit only to large scales  ($6-65$~{\Mpch}) as indicated by the vertical shaded regions.
    For reference, we show in grey the {\wgp} measurements of blue galaxies from KiDS$+$GAMA \citep{Johnston_2019} and 
    KiDS-1000 Bright$+$GAMA \citep{Georgiou2025} in the low-redshift panel (left) and
    DES$+$eBOSS ELG \citep{Samuroff_2023} in the middle panel.
    Bottom: the distributions of redshift and rest-frame absolute colour for the blue BGS samples, as well as the low and high redshift ELG samples. The colours match the corresponding measurements and the BGS parent distribution is shown in black.
    The contours correspond to the $39$ and $86$th percentiles.
    The average properties of the KiDS-1000 Bright 
    and eBOSS ELG 
    samples  are included for reference. 
    } 
    \label{fig:blue_low_ia}
\end{figure*}

\subsection{Modelling}
\label{sec:modelling}

We follow the modelling procedure described in Jeffrey et al. (in prep) and \cite{Samuroff_2023}.
We consider two models for intrinsic alignment: the Non-Linear linear-Alignment (NLA) model and the Tidal Alignment and Tidal Torquing model \citep[TATT;][]{Blazek2019}. 
The NLA model assumes the alignment of a galaxy is linearly proportional to the gravitational potential; see \cite{Lamman_2024} for a review. 
The amplitude of alignment is denoted {\AIA}.
The TATT model includes tidal alignment (linear in the tidal field), tidal torquing \citep[quadratic in the tidal field,][]{Mackey2002,Codis2015}, and source density weighting \citep{Blazek2015}.
At fixed redshift, the TATT model depends on a tidal alignment amplitude, $A_1$, a tidal torquing amplitude, $A_2$, and an effective source linear bias of the galaxies, $b_{\rm TA}$;
the NLA model is recovered by fixing $A_2=b_{\rm TA}=0$.

The models are evaluated using the CCL library \citep{Chisari2019}\footnote{\url{https://github.com/LSSTDESC/CCL}}. 
We jointly fit the shape--density {\wgp} and density--density {\wgg} correlations.
For NLA, we assume a linear galaxy bias model, and for TATT, we include second order galaxy bias.
This paper focuses on blue galaxy populations with weak alignment, therefore the effects of magnification and galaxy-galaxy lensing are neglected; these effects will be explored in the companion paper (Jeffrey et al., in prep).
For NLA, we limit the fit to large scales $6-65$~{\Mpch}, while for TATT, we extend the fit to $2-65$~{\Mpch}; in Appendix~\ref{sec:scale_dep_appendix}, we explore the dependence of the fits on the scale cuts (also see Jeffrey et al. in prep).
We adopt wide top-hat priors on the free parameters: $A_1$ (referred to as {\AIA} for NLA), $A_2$, $b_\mathrm{TA}$, $b_1$, and $b_2$.

To fully leverage the available data, we jointly model the IA measurements from the different weak lensing surveys; note, the measurements from the lensing surveys are statistically consistent (see Appendix~\ref{sec:survey_to_survey_appendix}).
For the BGS samples, we jointly model the measurements from the DES, KiDS, and SDSS shear catalogues;
to ensure independence between the measurements, we mask the SDSS shear catalogue to avoid overlap with DES and KiDS (Section~\ref{sec:sdss_imaging}). 
For the LRG samples, we jointly model the measurements from the DES and KiDS shear catalogues.
SDSS is omitted due to its limited imaging depth, which results in a bias towards low redshifts in the LRG--SDSS cross-matched shape catalogue (Figure~\ref{fig:samples_2d}).
HSC is omitted due to its significant overlap with the other lensing surveys.
For the ELG samples, HSC provides the highest signal-to-noise measurement, even compared to the DES--KiDS co-added measurements;  the depth of HSC imaging results in a high number of matches between the HSC shear catalogue and the ELG sample.
We therefore use the HSC cross-matched shear catalogue for our fiducial ELG measurements; 
our results are consistent with joint DES--KiDS modelling.

\begin{figure*}
    \includegraphics[width=\textwidth]{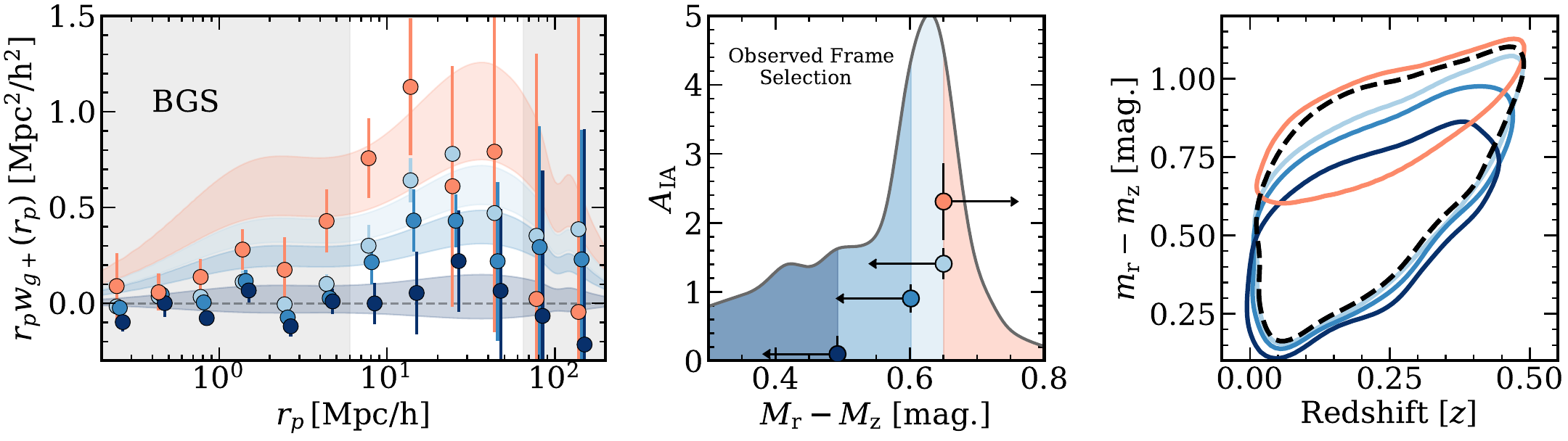}
    \caption{
    The non-detection of IA in blue galaxies depends on the purity of the blue sample, i.e., the amplitude of the IA signal scales with the rest-frame colour selection.
    We present the measured IA shape--density correlations in the left panel for three successively bluer BGS galaxy selections, defined as $M_\mathrm{r}-M_\mathrm{z}<$0.5 (as in Figure~\ref{fig:subsample_summary}), 0.6 and 0.65, in comparison to one red BGS selection ($M_\mathrm{r}-M_\mathrm{z} > 0.65$). The $16$th and $84$th percentiles of the NLA fit to each {\wgp} measurement are shown as shaded bands. The middle panel shows the distribution of the rest-frame colours for each sub-sample, where the y-axis indicates the amplitude of the NLA $A_\mathrm{IA}$ posteriors.  In all cases, the signals from DES, KiDS, and SDSS are jointly modelled. The right panel shows the four rest-frame colour selections in terms of apparent colour and redshift (only DES as a representative example), where the contours correspond to the $86$th percentile and the parent DES--BGS sample is shown in black for reference.
    } 
     \label{fig:blue_red_colorcuts}
\end{figure*}

\section{DESI IA: Characterising the Galaxy Population for Lensing 
Surveys}
\label{sec:characterizing_for_shear}

With DESI DR1 spectroscopy and shear catalogues from four lensing surveys, we present direct IA measurements for subsamples of the galaxy population (Section~\ref{sec:primary_measurements}). We fit these measurements with the NLA and TATT models and study the trends of the IA signal with galaxy properties to address three outstanding issues in the study of intrinsic alignments:
\begin{enumerate}
    \item Across redshift and scale, do blue galaxies align (Section~\ref{sec:blue_low_ia})?
    \item What is the optimal selection of the galaxy population for IA mitigation in cosmic shear analyses (Section~\ref{sec:blue_purity})?
    \item What are the physical drivers of IA in red, quenched galaxies (Section~\ref{sec:physical_drivers})?
\end{enumerate}

\subsection{IA measurements}
\label{sec:primary_measurements}

The IA shape--density correlation measurements for each DESI tracer are presented in Figure~\ref{fig:subsample_summary}. 
The detection significances of the measurements are reported in Table~\ref{tab:detection}. 
We include co-added shape--density measurements across the different lensing surveys, however, these co-added  measurements are not used for modelling; 
for the NLA and TATT modelling, we jointly fit the individual measurements. 

IA are significantly detected for the BGS sample $>10\sigma$. 
We divide each DESI BGS cross-matched shape catalogue into blue and red sub-populations by rest-frame colour, $M_r-M_z=0.5$, and then further divide by luminosity at $M_r=-20.9$ and $M_r=-21.9$ for the blue and red subsamples, respectively; 
the selection is only applied to the shape catalogues, and the density sample remains the complete BGS.  
The red BGS sample (upper row in Figure~\ref{fig:subsample_summary}) shows significantly stronger IA, while the blue BGS sample's IA signal is consistent with zero across all scales; we discuss this colour trend in Section~\ref{sec:blue_low_ia}.

The LRG measurements are significantly detected $>10\sigma$ and are presented in the middle panel of Figure~\ref{fig:subsample_summary}.
We divide the LRG shape catalogues first by redshift at $z=0.75$, followed by a luminosity split at $M_r=-23.1$ and $M_r=-23.3$ for the low and high redshift subsamples, respectively. 
The LRG correlation measurements exhibit strong colour and luminosity dependence (see Section~\ref{sec:physical_drivers}). 

The ELG IA measurements are consistent with zero (bottom panel of Figure~\ref{fig:subsample_summary}). 
This sample is divided into low and high redshift subsamples at $z=1.15$, and further by luminosity at $M_r=-21.3$ and $M_r=-21.9$, respectively. All ELG subsamples show no IA detection. These non-detections are discussed with the blue BGS galaxies in Section~\ref{sec:blue_low_ia}.

\subsection{Blue galaxy populations 
are consistent with no IA}
\label{sec:blue_low_ia}

IA has yet to be detected in samples of blue star-forming galaxies \citep{Mandelbaum2011,Johnston_2019,Samuroff_2023}.
Cosmic shear studies limited to blue galaxies --- \textit{blue shear} --- are therefore expected to be less biased by IA \citep{McCullough2024}. 
However, prior direct measurements of IA in blue galaxies were limited to $z \lesssim 0.8$ and report imprecise constraints on the alignment amplitude.
DESI~DR1's statistical power provides a more definitive constraint across a wider redshift range than previous IA studies of blue galaxies.

We report null detections of IA in the bluest galaxies between $z=0.05$~and~$1.55$. 
At low to moderate redshifts ($0.05<z<0.5$), we consider the magnitude limited BGS sample.
We define blue galaxies as those satisfying $M_\mathrm{r}-M_\mathrm{z}<0.5$ (Section~\ref{sec:data_samples});
the colour selection is only applied to the BGS shear catalogues, not the BGS density tracers.
The blue BGS galaxies are consistent with no shape--density correlations (reduced $\chi^2=0.20$ relative to a null signal);
below we discuss the sensitivity of the intrinsic alignment signal to the method of blue--red selection.
At higher redshifts ($0.8<z<1.55$), we consider the star forming ELG sample.
The ELGs are consistent with no intrinsic alignment, regardless of redshift.
These measurements are summarized in Figure~\ref{fig:blue_low_ia}, alongside prior measurements of blue galaxies from \cite{Johnston_2019,Samuroff_2023,Georgiou2025}.

To set informative priors on $A_{\rm IA}$ for cosmic shear, we model the measurements with NLA; see Section~\ref{sec:nla_v_tatt} for TATT modelling.
The {\wgg} and {\wgp} measurements are jointly fit between $6$ and $65$~{\Mpch}, assuming linear galaxy bias. For the blue BGS, low-$z$ ELG, and high-$z$ ELG measurements, we present the $1\sigma$ confidence intervals of the NLA fits in Figure~\ref{fig:blue_low_ia}.
The posteriors on the NLA parameters and goodness of fit are reported in Table~\ref{tab:NLA_TATT}.

Between the local Universe and $z \sim 1.6$, intrinsic alignments are undetected in the bluest galaxies.

\renewcommand{\arraystretch}{1.6}
\begin{table*}
\begin{center}
\begin{tabular}{cccclllc}

\hline
Sample & Rest-frame selection & Observed frame selection & Model & $A_1~(A_\mathrm{IA})$ & $A_2$ & $b_\mathrm{TA}$ & {\wgp} $\chi^2_\nu$\\
\hline
BGS & $M_r-M_z<0.5$  & $ m_\mathrm{r} - m_\mathrm{z} < 0.85 \times z + 0.55 $ & NLA &  $0.1_{-0.3}^{+0.3}$ & ---  & ---  & $1.0$ \\
 & &  & TATT &  $0.2_{-0.4}^{+0.4}$ & $-0.1_{-0.3}^{+0.6}$ & $-1.0_{-1.0}^{+1.0}$ & $2.13$ \\
\cline{2-8}
 & $M_r-M_z<0.6$  & $ m_\mathrm{r} - m_\mathrm{z} < 0.85 \times z + 0.65 $ & NLA &  $0.9_{-0.2}^{+0.2}$ & ---  & ---  & $0.93$ \\
 & &  & TATT &  $0.8_{-0.2}^{+0.3}$ & $0.8_{-0.5}^{+0.4}$ & $-1.5_{-0.3}^{+0.4}$ & $1.11$ \\
\cline{2-8}
 & $M_r-M_z<0.65$ & $ m_\mathrm{r} - m_\mathrm{z} < 0.85 \times z + 0.70 $ & NLA &  $1.4_{-0.2}^{+0.2}$ & ---  & ---  & $2.12$ \\
 & &  & TATT &  $1.4_{-0.3}^{+0.3}$ & $1.1_{-0.5}^{+0.5}$ & $-1.2_{-0.3}^{+0.2}$ & $1.9$ \\
\hline
ELG &  & Low-$z$: $z<1.15$ & NLA &  $0.0_{-1.0}^{+1.0}$ & ---  & ---  & $0.74$ \\
 & &  & TATT &  $-1.0_{-1.0}^{+2.0}$ & $3.0_{-4.0}^{+4.0}$ & $0.0_{-1.0}^{+1.0}$ & $3.04$ \\
\cline{2-8}
 &  & High-$z$: $z>1.15$ & NLA &  $-3.0_{-1.0}^{+1.0}$ & ---  & ---  & $1.18$ \\
 & &  & TATT &  $-2.0_{-3.0}^{+2.0}$ & $-1.0_{-5.0}^{+5.0}$ & $-1.0_{-1.0}^{+2.0}$ & $4.22$ \\
\hline
\end{tabular}
\caption{
Parameter posteriors and goodness of fit for NLA ($6-65$~{\Mpch}) and TATT ($2-65$~{\Mpch}) model fits to IA measurements.
Three selections of blue BGS galaxies, defined in terms of restframe colour $ M_\mathrm{r} - M_\mathrm{z}$, are presented. 
Approximations of the restframe selections in  observed frame properties ($m_\mathrm{r}-m_\mathrm{z}$ and $z$) are reported.
DES, KiDS, and SDSS are the shape tracers for the BGS measurements.
Two selections of ELG galaxies are also shown: low-$z$ ($0.8<z<1.15$) and high-$z$ ($1.15<z<1.55$).
HSC is the shape tracer for the ELG measurements.
An extended version of this table, including variations of the scale cuts, is presented in Appendix~\ref{sec:scale_dep_appendix}.
}
\label{tab:NLA_TATT}
\end{center}
\end{table*}

\subsection{On the purity of blue galaxy samples}
\label{sec:blue_purity}

\begin{figure}
    \includegraphics[width=0.9\columnwidth]{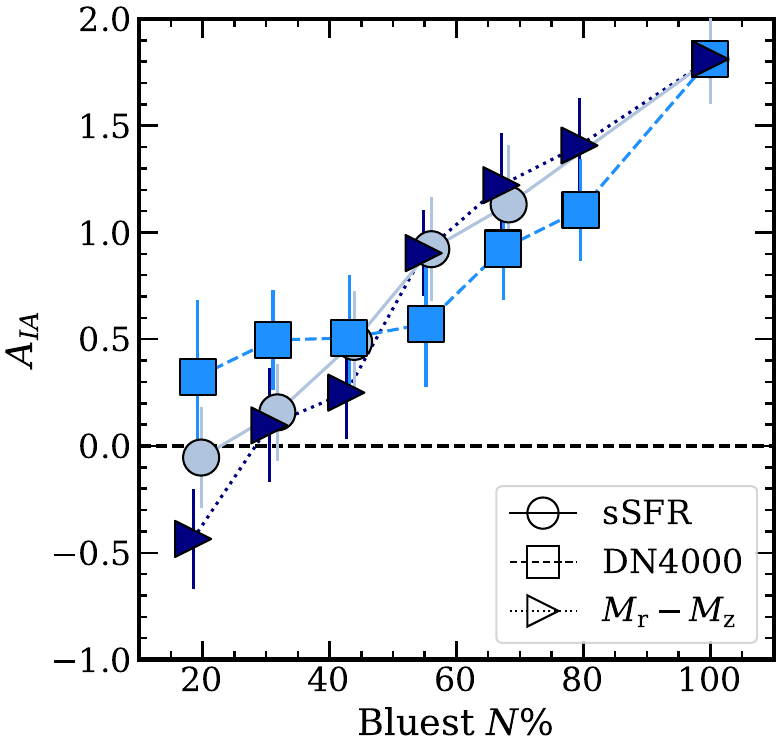}
    \caption{
    The posteriors on NLA $A_\mathrm{IA}$ for successively bluer galaxy samples.    
    DESI BGS is the density tracer and DES, KiDS, and SDSS are the shape tracers.
    We consider three definitions for selecting blue galaxies: specific star formation rate (sSFR), the strength of the $4000~\mathrm{\AA}$ break ({\DN}), and rest-frame colour. 
    For each delineator, we measure the intrinsic alignment signal of the \textit{bluest}  N\% of galaxies.
    } 
    \label{fig:blue_amplitudes}
\end{figure}

Limiting cosmic shear inference to the bluest galaxies comes at the cost of increased shot noise. Therefore, it is critical to optimise the selection of weakly aligned galaxies for future blue shear studies \citep{McCullough2024}.
How the alignment signal depends on galaxy colour also offers valuable insight into the nature of IA.

Using the magnitude limited BGS shear catalogues, we measure the intrinsic alignment signal for a series of increasingly blue galaxy populations, in terms of rest-frame colour $M_\mathrm{r}-M_\mathrm{z}$;
the cuts are only applied to the shear catalogues, not the density tracers.
Figure~\ref{fig:blue_red_colorcuts} presents the shape--density correlations for each colour selection; DESI BGS is the density tracer.
We model each measurement with NLA between $6$ and $65$~{\Mpch}, assuming linear galaxy bias.
The fits are presented alongside the measurements in Figure~\ref{fig:blue_red_colorcuts}.
The amplitude of intrinsic alignments {\AIA} declines for successively bluer galaxy samples until $M_\mathrm{r}-M_\mathrm{z}<0.50$ (approximately the bluest $30\%$), beyond which shape-density correlations are undetected at the $3\sigma$ level.

Rest-frame colour is one of several options for selecting blue galaxies.
To optimize the selection of galaxies for blue shear, we explore additional proxies of galaxy colour: strength of the $4000~\mathrm{\AA}$ break \citep[{\DN}, the ratio of continuum flux blueward and redward of the break,][]{Balogh1999} and specific star formation rate (star formation rate divided by stellar mass); both are derived from \texttt{fastspecfit}. 
For each colour tracer (rest-frame colour, {\DN}, and specific star formation rate), we measure the amplitude of intrinsic alignments with NLA for all galaxies bluer than the $N$th percentile (where $20 \leq N \leq 80$).
The alignment amplitudes {\AIA} are reported as a function of $N$ (i.e., severity of the selection) in Figure~\ref{fig:blue_amplitudes}.
The three proxies are equally efficient at selecting weakly aligned galaxies.

Throughout this paper, the rest-frame colour selection $M_\mathrm{r} - M_\mathrm{z} < 0.5$ is our fiducial definition of blue galaxies in the magnitude limited BGS shear catalogues.
To aid the selection of blue galaxies for cosmic shear, Table~\ref{tab:NLA_TATT} includes approximations of the rest-frame $M_\mathrm{r} - M_\mathrm{z}$ cuts in observed frame colour and redshift.

\subsection{IA model complexity: NLA vs TATT}
\label{sec:nla_v_tatt}

The NLA model neglects higher-order correlations, such as tidal torquing, which may be expected for blue spiral galaxies \citep[e.g.,][]{chisari_2015_hydrosim,des_y1_colorsplit}. 
We consider a five parameter TATT model to fit the measurements, assuming second order galaxy bias: $A_1$, $A_2$, $b_\mathrm{TA}$, $b_1$, and $b_2$; see Section~\ref{sec:modelling}.
We jointly fit {\wgg} and {\wgp} between $2$ and $65$~{\Mpch}.

\begin{figure*}
    \includegraphics[width=0.95\textwidth]{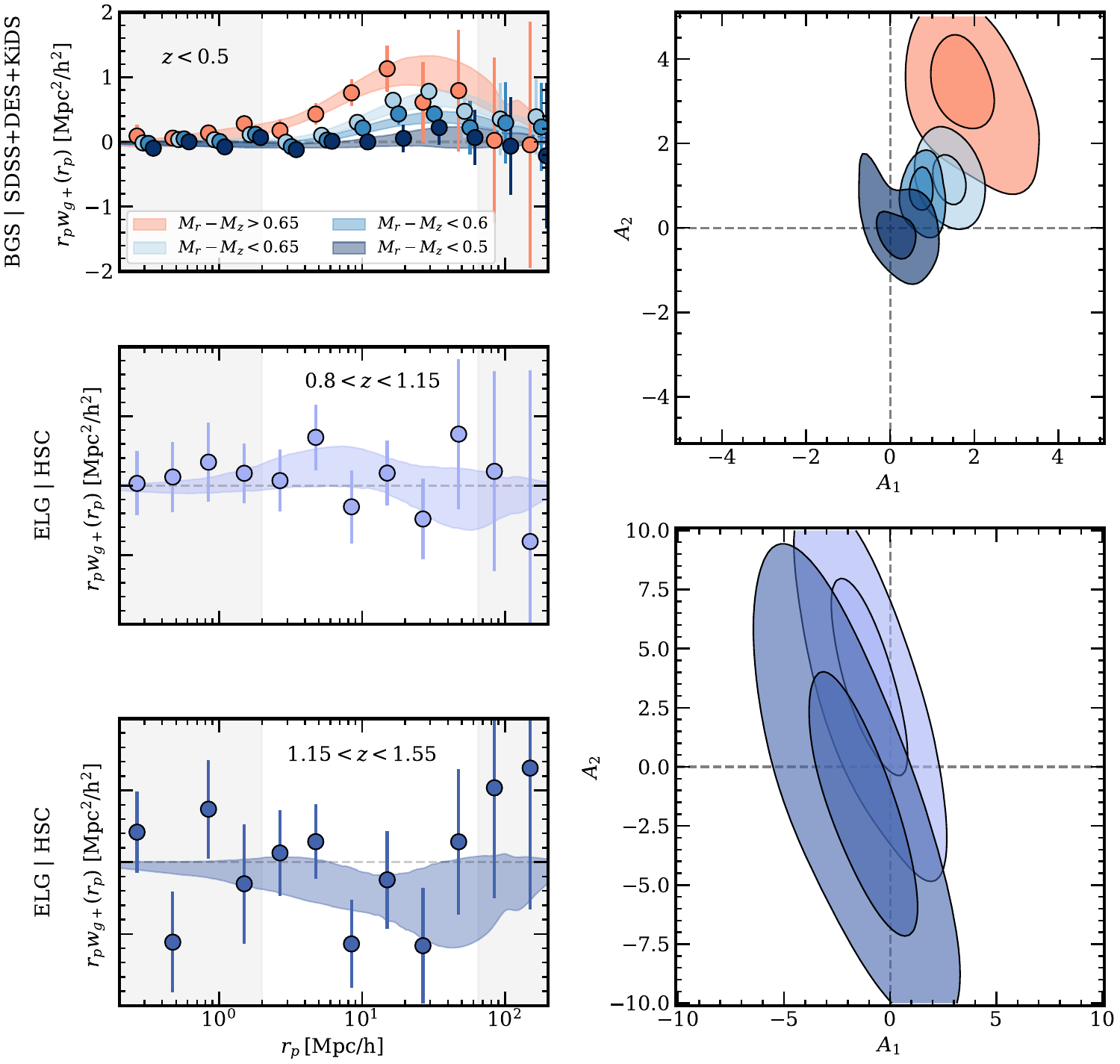}
    \caption{TATT fits and posteriors on $A_1,A_2$ for a series of successively bluer BGS galaxy samples ($0.05 < z<0.5$) as well as low-$z$ ($0.8 < z < 1.15$) and high-$z$ ($1.15 < z < 1.55$)  ELG samples; the BGS samples are identical to Figure~\ref{fig:blue_red_colorcuts}.
    The measured shape--density correlations are presented in the left-hand column, alongside the $1\sigma$ confidence intervals of the TATT fits.
    The right-hand column presents the posteriors on $A_1$ and $A_2$ for the BGS (top) and ELG  (bottom) measurements; contours correspond to the $39$ and $86$th percentiles.
    The parameter posteriors and goodness of fit are reported in Table~\ref{tab:NLA_TATT}.
    } 
    \label{fig:TATT}
\end{figure*}

For the bluest galaxy samples, the TATT alignment amplitudes $A_1$ and $A_2$ are consistent with zero. 
Figure~\ref{fig:TATT} presents the $1\sigma$ confidence intervals on the TATT models alongside the measured shape--density correlation, as well as the posteriors on $A_1$ and $A_2$. 
For more red contaminated samples (i.e., less severe blue selections), the TATT model fits the data well at all scales; see Jeffrey et al. (in prep) for more discussion of extending the TATT model to small scales.
Table~\ref{tab:NLA_TATT} reports the parameter posteriors and goodness of fit for the models.

\section{Physical Drivers of Intrinsic Alignment}
\label{sec:physical_drivers}

Priors studies have begun to map the nature of intrinsic alignments across galaxy property space.
Alignments are found to be stronger in brighter and redder elliptical samples \citep{Mandelbaum2006, Hirata2007,Joachimi2011,Singh2015,Johnston_2019,Fortuna2021,Samuroff_2023,HervasPeters2024, Georgiou2025,CristinaFortuna2025}, and no significant dependence on redshift has been measured \citep[e.g.,][]{Fortuna2021, paus_ia}. 
Inferring these trends from a patchwork of different lensing and spectroscopic surveys is difficult. 
Previous studies probed narrow ranges of redshift and colour.
Galaxy properties are also highly correlated, making it difficult to separate the contributions of galaxy colour, luminosity, and redshift. 
With large samples spanning a wide range of galaxy types, DESI~DR1 is well poised to fill in the picture of intrinsic alignments.

\begin{figure}
    \includegraphics[width=\columnwidth]{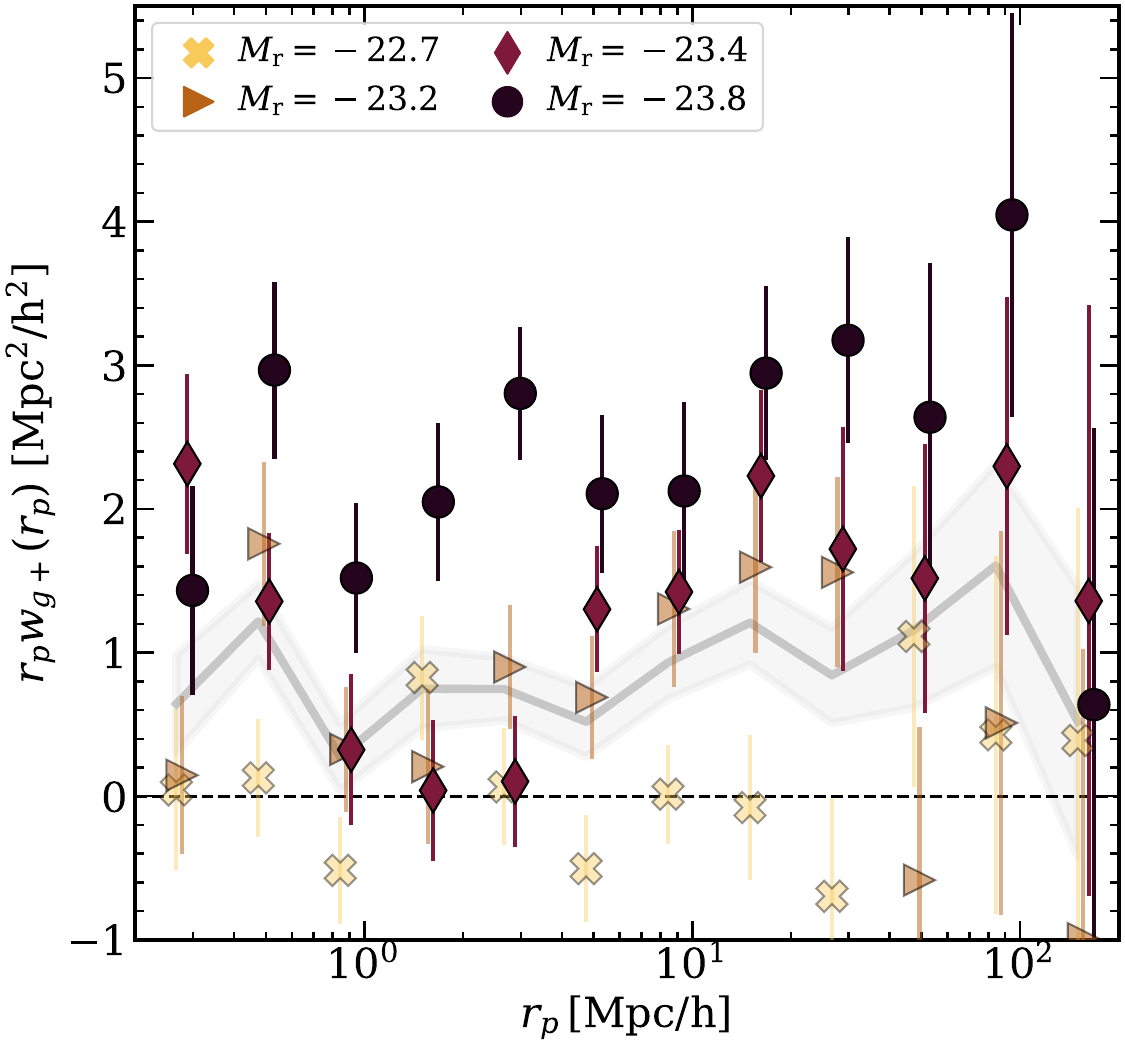}
    \caption{ 
    For LRGs, the amplitude of the intrinsic alignment signal increases with galaxy luminosity.
    With DESI LRG as the density tracer and DES$+$KiDS as the shape tracer, we measure the IA correlation in four luminosity bins.
    The average luminosity of each bin is reported.
    The IA signal from the parent LRG sample (i.e., no cuts) is shown in grey.
    } 
    \label{fig:lrg_luminosity}
\end{figure}

\begin{figure*}

    \includegraphics[width=\textwidth]{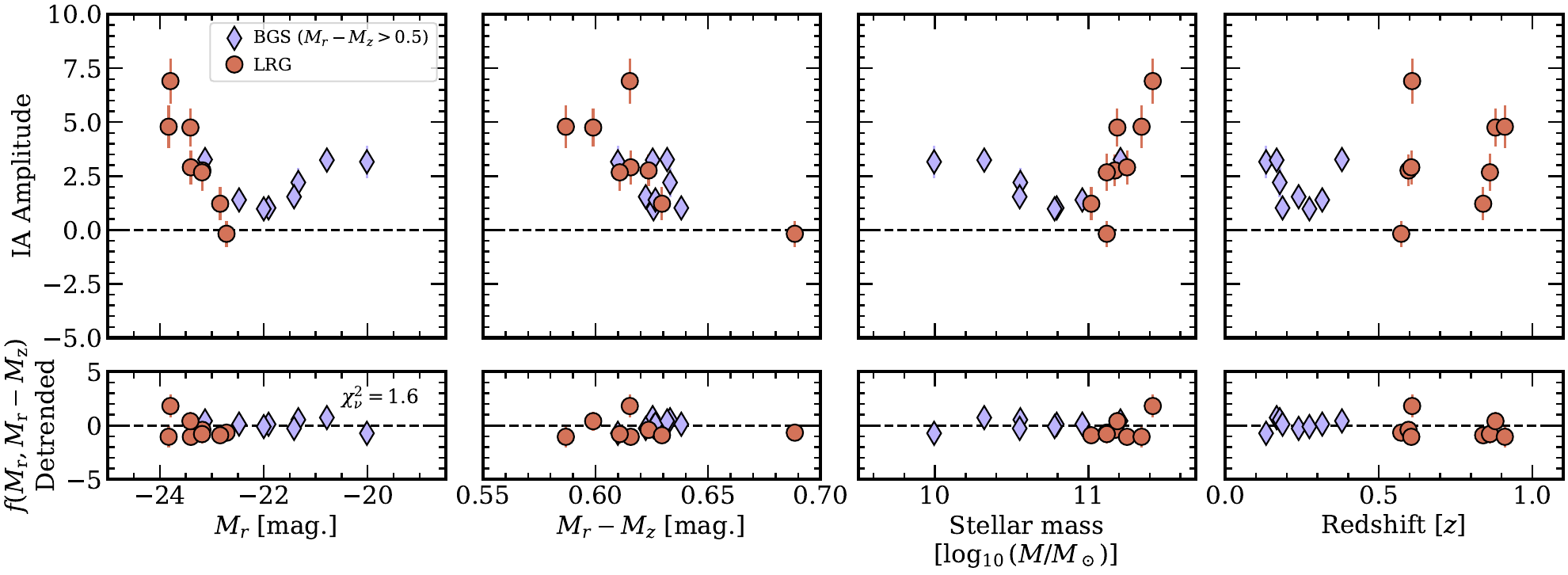}
    \caption{ 
   An investigation of model free IA amplitude (Equation~\ref{eqn:amplitude}) in mass, luminosity, colour, and redshift space.  
   We consider the BGS and LRG samples.
   The BGS matched DES, KiDS, and SDSS shape catalogues are first limited to red ($M_\mathrm{r} - M_\mathrm{z}>0.5$) populations, each is then divided into two redshift bins, and again into four luminosity bins (totalling 8 samples).
   We divide the LRG shape catalogues into two redshift ranges---low ($0.4 < z < 0.75$) and high ($0.75 < z < 1.10$)---and each into four luminosity bins.
   DESI BGS (LRG) is used as the density tracer for the BGS (LRG) matched shape catalogues.
   Top row: the amplitude of intrinsic alignments as a function of mean  luminosity, rest-frame colour, stellar mass, and redshift.
   Bottom row: the residual intrinsic alignment amplitudes after subtracting the best-fit amplitude--luminosity--colour relation (described in Section~\ref{sec:physical_drivers}). 
    } 
    \label{fig:galaxy_property_mega}
\end{figure*}

Here we investigate the dependence of red galaxy IA on luminosity, mass, colour, and redshift.
To aid our investigation, we define a model-free intrinsic alignment amplitude:
\begin{equation}
\label{eqn:amplitude}
\mathrm{Amplitude} = M \times \frac{\sum_{ r_\mathrm{min} < r_{\rm p} < r_\mathrm{max} } w_{\rm g+}(r_{\rm p})}{\sum_{ r_\mathrm{min} < r_{\rm p} < r_\mathrm{max}} \sqrt{w_{\rm gg}(r_{\rm p})}} + B.
\end{equation}
We set $r_\mathrm{min},r_\mathrm{max} = 6$ and $65~\Mpch$ to restrict the measurement to linear scales, where we can neglect higher-order bias terms (analogous to NLA). 
The coefficients $M=106,B=0.3$ scale the model-free amplitude to approximate $A_{\rm IA}$. 
We reserve more complex modelling of the red galaxy samples for Jeffrey et al. (in prep).

We first consider the LRG sample of red quenched galaxies.
The LRG-matched shear catalogues are each divided into four luminosity bins;
the bins are defined in terms of percentiles of $M_r$, with bin edges of $[0,20,40,60,100]$.
Figure~\ref{fig:lrg_luminosity} presents the measured IA signal of each luminosity bin. 
Consistent with prior studies, the brighter samples display greater alignment. 
The lowest luminosity bin $\langle M_\mathrm{r} \rangle = -22.7$ is consistent with zero alignment.
These measurements should not be interpreted in terms of luminosity alone. 
Due to correlations within the LRG sample, the luminosity bins each probe different rest-frame colours; in the LRG sample, brighter galaxies are bluer (Figure~\ref{fig:samples_2d}). 
Intrinsic alignment trends must therefore be jointly studied in luminosity, mass, colour, and redshift space.

To further probe the alignment of galaxies, we divide the LRG- and BGS-matched shear catalogues into narrow bins in luminosity, colour, and redshift space.
The ELG-matched shear catalogues are omitted from this study, because no alignment signal is detected.
We first divide the LRG-matched shear catalogues into eight independent samples: we consider the low ($0.4<z<0.75$) and high ($0.75<z<1.1$) redshift bins introduced in Section~\ref{sec:data_samples} and divide each into four luminosity bins;
as above, the luminosity bins are defined in terms of percentiles of $M_r$, with bin edges of $[0,20,40,60,100]$.
The model free amplitude of the IA signal for each bin is presented in Figure~\ref{fig:galaxy_property_mega}.
The shape--density correlations are measured relative to the LRG density tracers.
We follow a similar binning scheme for the BGS-matched shear catalogues, dividing the red BGS sample ($M_\mathrm{r} - M_\mathrm{z} > 0.5$) into 8 bins:
the red BGS population is first divided into low ($0.05<z<0.2$) and high ($0.2<z<0.5$) redshift bins, each sample is then cut into four luminosity bins to achieve equal number of objects in each.
The intrinsic alignment signal is measured for each sub-sample relative to the BGS density tracer catalogue.
The model free IA amplitudes are presented alongside the LRG measurements in Figure~\ref{fig:galaxy_property_mega}.

With the BGS and LRG samples defined above, our IA measurements span four magnitudes in luminosity and reach from the local Universe to $z \sim 1$.
The relationship between IA amplitude and a particular galaxy property is complicated by the correlations between luminosity, rest-frame colour, mass, and redshift (e.g., a correlation between amplitude and redshift could result from a redshift--luminosity correlation).
To identify the primary drivers of IA, we therefore model the red samples as a function of galaxy properties.
We limit our model to red galaxies for simplicity;
however, given the lack of alignments in blue galaxies, a model of red galaxy alignment can be approximately extended to all colours by setting the amplitude to zero below a certain rest-frame colour.
The red samples are well described as a function of luminosity and rest-frame colour: we consider the sum of a double power-law in luminosity \citep{Fortuna2021} and a linear function in rest-frame colour. 
The model yields a reduced $\chi^2_\nu=1.6$;
a double power-law in luminosity yields $\chi^2_\nu=2$ and a linear function of colour yields $\chi^2_\nu=6$.
Figure~\ref{fig:galaxy_property_mega} presents the model residuals, revealing no lingering trends in luminosity, colour, mass, or redshift.
Further modelling will be presented in Jeffrey et al. (in prep).

With DESI~DR1, we surveyed the amplitude of IA across a wide swath of galaxy property space. 
The measurements are well described by a function of luminosity and rest-frame colour, with no residual redshift or stellar mass dependence.

\begin{figure*}
    \includegraphics[width=\textwidth]
    {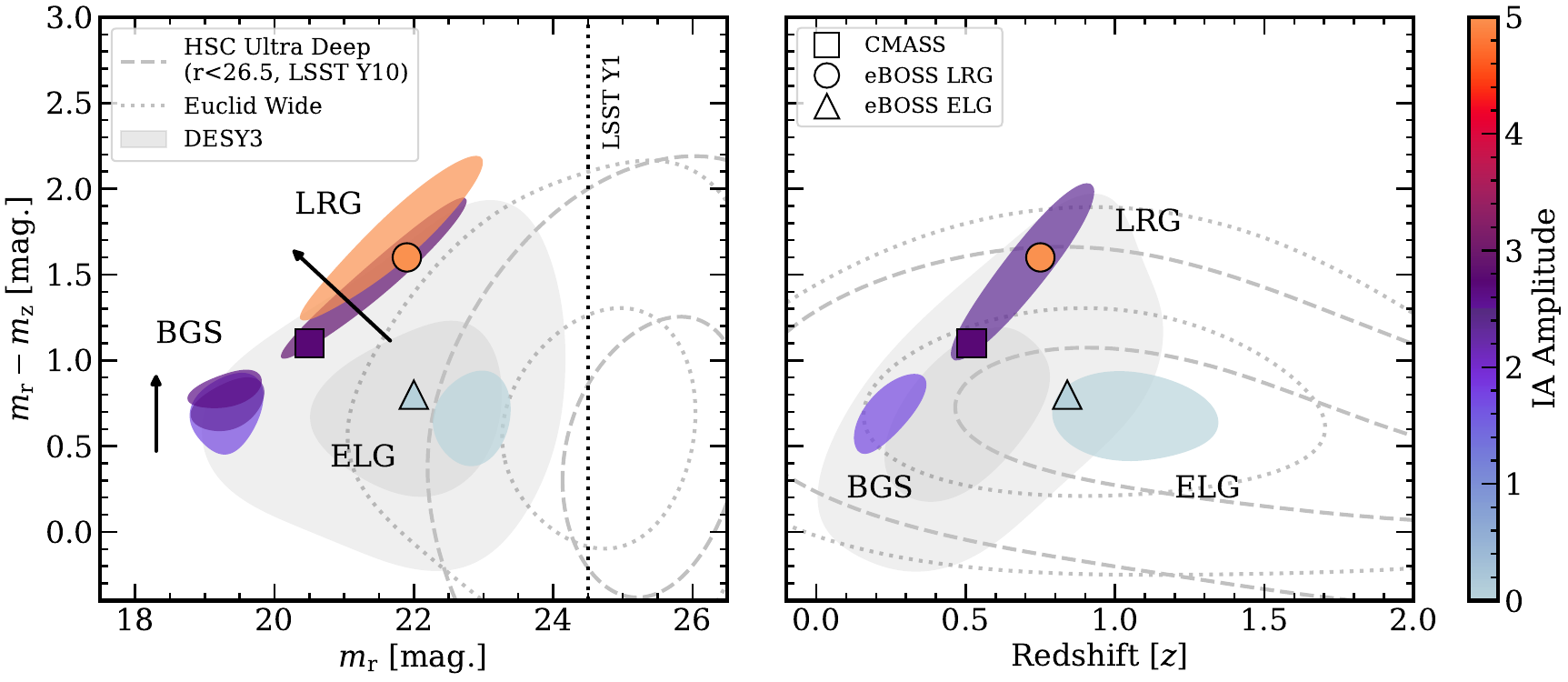}
    \caption{ 
    The landscape of direct intrinsic alignment measurements.
    We present a representative stage III lensing survey (DES~Y3, grey) and the DESI DR1 cross-matched samples (BGS, LRG, ELG), in apparent magnitude--colour (left) and redshift--colour space (right).
    The DES~Y3 contours correspond to the $39$ and $86$th percentiles; for the DESI samples, only the $39$th percentiles are shown for clarity.
    On the left, trends in intrinsic alignment are visualized by dividing the BGS and LRG cross-matched samples in terms of rest-frame colour and luminosity, respectively.
    The arrows indicate regions of greater alignment amplitude.
    Prior measurements \citep{Samuroff_2023} are included for reference.
    To forecast a LSST-like lensing survey, the HSC Ultra Deep photometric catalogue is shown in dashed-grey contours (limited to a LSST~Y10 depth of $r<26.5$).
    The depth of LSST~Y1 $r<24.5$ is demarcated by a vertical dotted line.
    The Euclid Q1 source catalogue, representative of the Wide Survey, is shown in dotted-grey contours.
    } 
    \label{fig:survey_magnitudes}
\end{figure*}

\section{Landscape of measurements}
\label{sec:landscape}
Precise direct measurements of intrinsic alignment require a dense spectroscopic sample over a contiguous area with overlapping shear measurements---increasingly possible with the advent of massively multiplexed spectroscopic instruments (e.g., DESI, \citealt{DESIinstrument2022}; PFS, \citealt{pfs_latest}; 4MOST, \citealt{4MOST}). 
Since IA is known to depend on galaxy colour and luminosity, it is important to measure IA for galaxy samples representative of weak gravitational lensing surveys.
This requires understanding the selection functions of the spectroscopic and imaging surveys. 

We summarize the current landscape of direct measurements from DESI~DR1 in Figure~\ref{fig:survey_magnitudes}.
For context, we include the DES~Y3, Euclid Q1,\footnote{\url{https://www.cosmos.esa.int/web/euclid/q1-data}} and HSC Ultra-Deep  source catalogues; a magnitude cut is applied to the latter catalogue to mock the expected ten-year depth of the Legacy Survey of Space and Time (LSST). The DESI tracers expand the parameter space covered by previous measurements \citep[e.g.,][]{Samuroff_2023} and bookend the colour--magnitude space of the Stage III lensing populations; however, the DESI samples still fall short of characterising the faint galaxy populations expected from LSST. 
Deeper spectroscopy, high-quality photometric redshifts, and indirect measurements will be necessary to perform IA measurements for future weak-lensing surveys.

In the next decade, DESI-II \citep[][]{snowmass_desi2} aims to observe $\sim$40 million galaxies.
DESI~II will observed at higher density in the $z < 1$ Universe (particularly relevant for IA) and at higher redshifts ($z >2$), compared to DESI. 
The 4-metre Multi Object Spectroscopic Telescope (4MOST; \citealt{4MOST})  will cover the southern sky with $\sim$2400 fibres per exposure, observing a wide variety of target classes\footnote{\url{https://www.4most.eu/cms/science/overview/}} with optimal overlap for LSST. 
These next-generation data sets will be powerful probes of IA, essential for Stage IV cosmology.

\section{Conclusions}
\label{sec:conclusions}

We present direct measurements of intrinsic alignments with the DESI~DR1 spectroscopic sample and four lensing surveys: DES, KiDS, HSC, and SDSS.
Our galaxy sample includes over $2$~million galaxies with precise redshift and shape measurements between the local Universe and $z\sim1.5$.
This sample is more than ten times the size of prior spectroscopic direct intrinsic alignment measurements and spans a colour, luminosity, and redshift space that is relevant to modern lensing shape catalogues. 
For the BGS and LRG samples, we detect alignments with $>10\sigma$ significance; no alignments are detected for the ELG samples.
We report informative priors for cosmic shear from NLA and TATT modelling.

\begin{enumerate}
\item \textbf{We present the most comprehensive library of direct IA measurements to date}, with over 20
independent subsamples of red and blue galaxies, across four magnitudes in luminosity and $z=0-1.5$
(Figure~\ref{fig:galaxy_property_mega}). 
Alignment is found to be strongest in redder and brighter galaxy samples.
We find alignment amplitude is entirely explained in terms of luminosity and colour trends alone, with no need for additional redshift dependence.

\item \textbf{Intrinsic alignments are not detected in blue star forming galaxies} between the local Universe and $z=1.5$.
The NLA and TATT alignment amplitudes are consistent with zero.
These measurements significantly extend the redshift range at which alignments are undetected in blue galaxies and lower the upper limit on alignments in blue galaxies at low redshift (Figure~\ref{fig:blue_low_ia}). 

\item \textbf{We report an optimal selection of blue unaligned galaxies.} 
Multiple tracers of stellar age---rest-frame colour, the strength of the $4000~\mathrm{\AA}$ break ({\DN}), and specific star formation rate---are equally correlated with intrinsic alignment amplitude.
For ease of use, we report our optimal selection in terms of redshift and observed frame $r-z$ colour (Figure~\ref{fig:blue_red_colorcuts} and Table~\ref{tab:NLA_TATT}).

\end{enumerate}

\textbf{Outlook:} Our measurements will inform the next generation of intrinsic alignment models for cosmic shear and the optimal selection of unaligned galaxies (McCullough et al. in prep.). 
With DESI~DR1, direct measurements of intrinsic alignment now span most of the colour, luminosity, and redshift space of modern shear catalogues.
In the near future, new spectroscopic samples \citep[e.g., Euclid and 4MOST,][]{Laureijs2011,4MOST} and next generation lensing surveys \citep[e.g., Vera Rubin, Euclid, and Roman,][]{Ivezic2019,Laureijs2011,Akeson2019} will continue to fill in the picture of intrinsic alignments and push our measurements to higher redshifts.

\bibliographystyle{mnras}
\bibliography{sources}

\section*{Affiliations}
\small
$^{1}$ Department of Astrophysical Sciences, Princeton University, Princeton NJ 08544, USA\\
$^{2}$ Kavli Institute for Cosmology, University of Cambridge, Madingley Road, Cambridge CB3 0HA, UK\\
$^{3}$ Center for Astrophysics $|$ Harvard \& Smithsonian, 60 Garden Street, Cambridge, MA 02138, USA\\
$^{4}$ Department of Physics \& Astronomy, University College London, Gower Street, London, WC1E 6BT, UK\\
$^{5}$ Leiden Observatory, Leiden University, Niels Bohrweg 2, 2333 CA, Leiden, the Netherlands\\
$^{6}$ Department of Astronomy and Astrophysics, UCO/Lick Observatory, University of California, 1156 High Street, Santa Cruz, CA 95064, USA\\
$^{7}$ Center for Cosmology and AstroParticle Physics, The Ohio State University, 191 West Woodruff Avenue, Columbus, OH 43210, USA\\
$^{8}$ Department of Astronomy, The Ohio State University, 4055 McPherson Laboratory, 140 W 18th Avenue, Columbus, OH 43210, USA\\
$^{9}$ The Ohio State University, Columbus, 43210 OH, USA\\
$^{10}$ Lawrence Berkeley National Laboratory, 1 Cyclotron Road, Berkeley, CA 94720, USA\\
$^{11}$ Department of Physics, Boston University, 590 Commonwealth Avenue, Boston, MA 02215 USA\\
$^{12}$ Dipartimento di Fisica ``Aldo Pontremoli'', Universit\`a degli Studi di Milano, Via Celoria 16, I-20133 Milano, Italy\\
$^{13}$ INAF-Osservatorio Astronomico di Brera, Via Brera 28, 20122 Milano, Italy\\
$^{14}$ Centre for Astrophysics \& Supercomputing, Swinburne University of Technology, P.O. Box 218, Hawthorn, VIC 3122, Australia\\
$^{15}$ Institut d'Estudis Espacials de Catalunya (IEEC), c/ Esteve Terradas 1, Edifici RDIT, Campus PMT-UPC, 08860 Castelldefels, Spain\\
$^{16}$ Institute of Space Sciences, ICE-CSIC, Campus UAB, Carrer de Can Magrans s/n, 08913 Bellaterra, Barcelona, Spain\\
$^{17}$ Instituto de F\'{\i}sica, Universidad Nacional Aut\'{o}noma de M\'{e}xico,  Circuito de la Investigaci\'{o}n Cient\'{\i}fica, Ciudad Universitaria, Cd. de M\'{e}xico  C.~P.~04510,  M\'{e}xico\\
$^{18}$ Physics Department, Brookhaven National Laboratory, Upton, NY 11973, USA\\
$^{19}$ University of California, Berkeley, 110 Sproul Hall \#5800 Berkeley, CA 94720, USA\\
$^{20}$ Institut de F\'{i}sica d’Altes Energies (IFAE), The Barcelona Institute of Science and Technology, Edifici Cn, Campus UAB, 08193, Bellaterra (Barcelona), Spain\\
$^{21}$ Departamento de F\'isica, Universidad de los Andes, Cra. 1 No. 18A-10, Edificio Ip, CP 111711, Bogot\'a, Colombia\\
$^{22}$ Observatorio Astron\'omico, Universidad de los Andes, Cra. 1 No. 18A-10, Edificio H, CP 111711 Bogot\'a, Colombia\\
$^{23}$ Institute of Cosmology and Gravitation, University of Portsmouth, Dennis Sciama Building, Portsmouth, PO1 3FX, UK\\
$^{24}$ University of Virginia, Department of Astronomy, Charlottesville, VA 22904, USA\\
$^{25}$ Fermi National Accelerator Laboratory, PO Box 500, Batavia, IL 60510, USA\\
$^{26}$ Department of Physics, The Ohio State University, 191 West Woodruff Avenue, Columbus, OH 43210, USA\\
$^{27}$ Department of Physics, The University of Texas at Dallas, 800 W. Campbell Rd., Richardson, TX 75080, USA\\
$^{28}$ CIEMAT, Avenida Complutense 40, E-28040 Madrid, Spain\\
$^{29}$ Department of Physics, Southern Methodist University, 3215 Daniel Avenue, Dallas, TX 75275, USA\\
$^{30}$ Department of Physics and Astronomy, University of California, Irvine, 92697, USA\\
$^{31}$ Perimeter Institute for Theoretical Physics, 31 Caroline St. North, Waterloo, ON N2L 2Y5, Canada\\
$^{32}$ Sorbonne Universit\'{e}, CNRS/IN2P3, Laboratoire de Physique Nucl\'{e}aire et de Hautes Energies (LPNHE), FR-75005 Paris, France\\
$^{33}$ Departament de F\'{i}sica, Serra H\'{u}nter, Universitat Aut\`{o}noma de Barcelona, 08193 Bellaterra (Barcelona), Spain\\
$^{34}$ NSF NOIRLab, 950 N. Cherry Ave., Tucson, AZ 85719, USA\\
$^{35}$ Instituci\'{o} Catalana de Recerca i Estudis Avan\c{c}ats, Passeig de Llu\'{\i}s Companys, 23, 08010 Barcelona, Spain\\
$^{36}$ Department of Physics and Astronomy, Siena College, 515 Loudon Road, Loudonville, NY 12211, USA\\
$^{37}$ Department of Physics \& Astronomy and Pittsburgh Particle Physics, Astrophysics, and Cosmology Center (PITT PACC), University of Pittsburgh, 3941 O'Hara Street, Pittsburgh, PA 15260, USA\\
$^{38}$ Departamento de F\'{\i}sica, DCI-Campus Le\'{o}n, Universidad de Guanajuato, Loma del Bosque 103, Le\'{o}n, Guanajuato C.~P.~37150, M\'{e}xico\\
$^{39}$ Instituto Avanzado de Cosmolog\'{\i}a A.~C., San Marcos 11 - Atenas 202. Magdalena Contreras. Ciudad de M\'{e}xico C.~P.~10720, M\'{e}xico\\
$^{40}$ IRFU, CEA, Universit\'{e} Paris-Saclay, F-91191 Gif-sur-Yvette, France\\
$^{41}$ Department of Physics and Astronomy, University of Waterloo, 200 University Ave W, Waterloo, ON N2L 3G1, Canada\\
$^{42}$ Waterloo Centre for Astrophysics, University of Waterloo, 200 University Ave W, Waterloo, ON N2L 3G1, Canada\\
$^{43}$ Instituto de Astrof\'{i}sica de Andaluc\'{i}a (CSIC), Glorieta de la Astronom\'{i}a, s/n, E-18008 Granada, Spain\\
$^{44}$ Departament de F\'isica, EEBE, Universitat Polit\`ecnica de Catalunya, c/Eduard Maristany 10, 08930 Barcelona, Spain\\
$^{45}$ Department of Physics and Astronomy, Sejong University, 209 Neungdong-ro, Gwangjin-gu, Seoul 05006, Republic of Korea\\
$^{46}$ Max Planck Institute for Extraterrestrial Physics, Gie\ss enbachstra\ss e 1, 85748 Garching, Germany\\
$^{47}$ Department of Physics, University of Michigan, 450 Church Street, Ann Arbor, MI 48109, USA\\
$^{48}$ University of Michigan, 500 S. State Street, Ann Arbor, MI 48109, USA\\
$^{49}$ Department of Astronomy, Tsinghua University, 30 Shuangqing Road, Haidian District, Beijing, China, 100190\\
$^{50}$ National Astronomical Observatories, Chinese Academy of Sciences, A20 Datun Road, Chaoyang District, Beijing, 100101, P.~R.~China\\
\normalsize

\section*{Data Availability}
The spectroscopic and imaging data in this paper are publicly available (for DESI spectroscopic redshifts and value-added catalogues see \citealt{DESIDR12025})\footnote{\url{https://data.desi.lbl.gov/doc/releases/dr1/}}. The  {\wgg}  measurements, {\wgp} measurements, and $N(z)$ distributions for the matched redshift--shape catalogues will be made available upon publication of the companion modelling paper  (Jeffreys et al., in prep.) on the echo-IA repository.\footnote{\url{https://github.com/echo-IA/IAmeasurementsStore}}

\section*{Acknowledgments}

JS acknowledges support by the National Science Foundation Graduate Research Fellowship Program under Grant DGE-2039656. 
Any opinions, findings, and conclusions or recommendations expressed in this material are those of the author(s) and do not necessarily reflect the views of the National Science Foundation.
NJ and BJ are supported by the ERC-selected UKRI Frontier Research Grant EP/Y03015X/1 and by STFC Consolidated Grant ST/V000780/1.

\textbf{DESI: }This material is based upon work supported by the U.S. Department of Energy (DOE), Office of Science, Office of High-Energy Physics, under Contract No. DE–AC02–05CH11231, and by the National Energy Research Scientific Computing Center, a DOE Office of Science User Facility under the same contract. Additional support for DESI was provided by the U.S. National Science Foundation (NSF), Division of Astronomical Sciences under Contract No. AST-0950945 to the NSF’s National Optical-Infrared Astronomy Research Laboratory; the Science and Technology Facilities Council of the United Kingdom; the Gordon and Betty Moore Foundation; the Heising-Simons Foundation; the French Alternative Energies and Atomic Energy Commission (CEA); the National Council of Humanities, Science and Technology of Mexico (CONAHCYT); the Ministry of Science, Innovation and Universities of Spain (MICIU/AEI/10.13039/501100011033), and by the DESI Member Institutions: \url{https://www.desi.lbl.gov/collaborating-institutions}. Any opinions, findings, and conclusions or recommendations expressed in this material are those of the author(s) and do not necessarily reflect the views of the U. S. National Science Foundation, the U. S. Department of Energy, or any of the listed funding agencies.

The authors are honored to be permitted to conduct scientific research on I'oligam Du'ag (Kitt Peak), a mountain with particular significance to the Tohono O’odham Nation.

\vspace{0.3cm}
\noindent\textbf{Imaging surveys: }

\textit{KiDS-1000: }Based on observations made with ESO Telescopes at the La Silla Paranal Observatory under programme IDs 177.A-3016, 177.A-3017, 177.A-3018 and 179.A-2004, and on data products produced by the KiDS consortium. The KiDS production team acknowledges support from: Deutsche Forschungsgemeinschaft, ERC, NOVA and NWO-M grants; Target; the University of Padova, and the University Federico II (Naples).

\textit{DES Y3: }This project used public archival data from the Dark Energy Survey (DES). Funding for the DES Projects has been provided by the U.S. Department of Energy, the U.S. National Science Foundation, the Ministry of Science and Education of Spain, the Science and Technology FacilitiesCouncil of the United Kingdom, the Higher Education Funding Council for England, the National Center for Supercomputing Applications at the University of Illinois at Urbana-Champaign, the Kavli Institute of Cosmological Physics at the University of Chicago, the Center for Cosmology and Astro-Particle Physics at the Ohio State University, the Mitchell Institute for Fundamental Physics and Astronomy at Texas A\&M University, Financiadora de Estudos e Projetos, Funda{\c c}{\~a}o Carlos Chagas Filho de Amparo {\`a} Pesquisa do Estado do Rio de Janeiro, Conselho Nacional de Desenvolvimento Cient{\'i}fico e Tecnol{\'o}gico and the Minist{\'e}rio da Ci{\^e}ncia, Tecnologia e Inova{\c c}{\~a}o, the Deutsche Forschungsgemeinschaft, and the Collaborating Institutions in the Dark Energy Survey.
The Collaborating Institutions are Argonne National Laboratory, the University of California at Santa Cruz, the University of Cambridge, Centro de Investigaciones Energ{\'e}ticas, Medioambientales y Tecnol{\'o}gicas-Madrid, the University of Chicago, University College London, the DES-Brazil Consortium, the University of Edinburgh, the Eidgen{\"o}ssische Technische Hochschule (ETH) Z{\"u}rich,  Fermi National Accelerator Laboratory, the University of Illinois at Urbana-Champaign, the Institut de Ci{\`e}ncies de l'Espai (IEEC/CSIC), the Institut de F{\'i}sica d'Altes Energies, Lawrence Berkeley National Laboratory, the Ludwig-Maximilians Universit{\"a}t M{\"u}nchen and the associated Excellence Cluster Universe, the University of Michigan, the National Optical Astronomy Observatory, the University of Nottingham, The Ohio State University, the OzDES Membership Consortium, the University of Pennsylvania, the University of Portsmouth, SLAC National Accelerator Laboratory, Stanford University, the University of Sussex, and Texas A\&M University.
Based in part on observations at Cerro Tololo Inter-American Observatory, National Optical Astronomy Observatory, which is operated by the Association of Universities for Research in Astronomy (AURA) under a cooperative agreement with the National Science Foundation.

\appendix

\section{DESI--Lensing Cross-matched Samples}
\label{sec:crossmatching_appendix}

\begin{figure*}
\includegraphics[width=\textwidth]{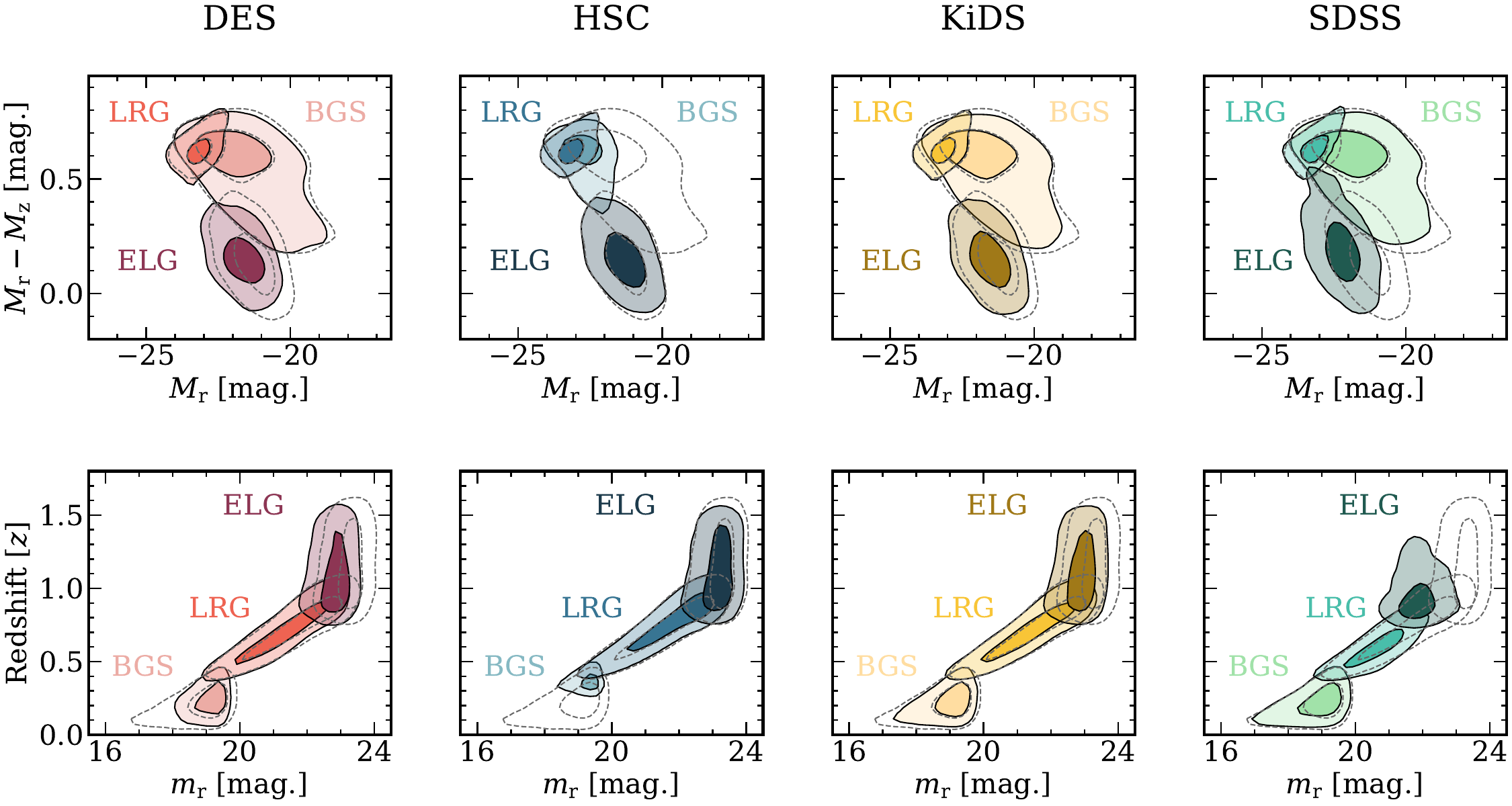}
\caption{
The properties of the DESI DR1 BGS, LRG, and ELG  galaxy samples before (grey dashed) and after cross-matching with DES (red), HSC (blue),  KiDS (yellow), and SDSS (green) imaging.
Top row: the galaxies' rest-frame absolute magnitudes and colours. 
Bottom row: the galaxies' apparent magnitudes and redshifts.
The contours indicate the $39$ and $86$th percentiles. 
} 
\label{fig:samples_2d}
\end{figure*}

\begin{figure*}
    \includegraphics[width=\textwidth]{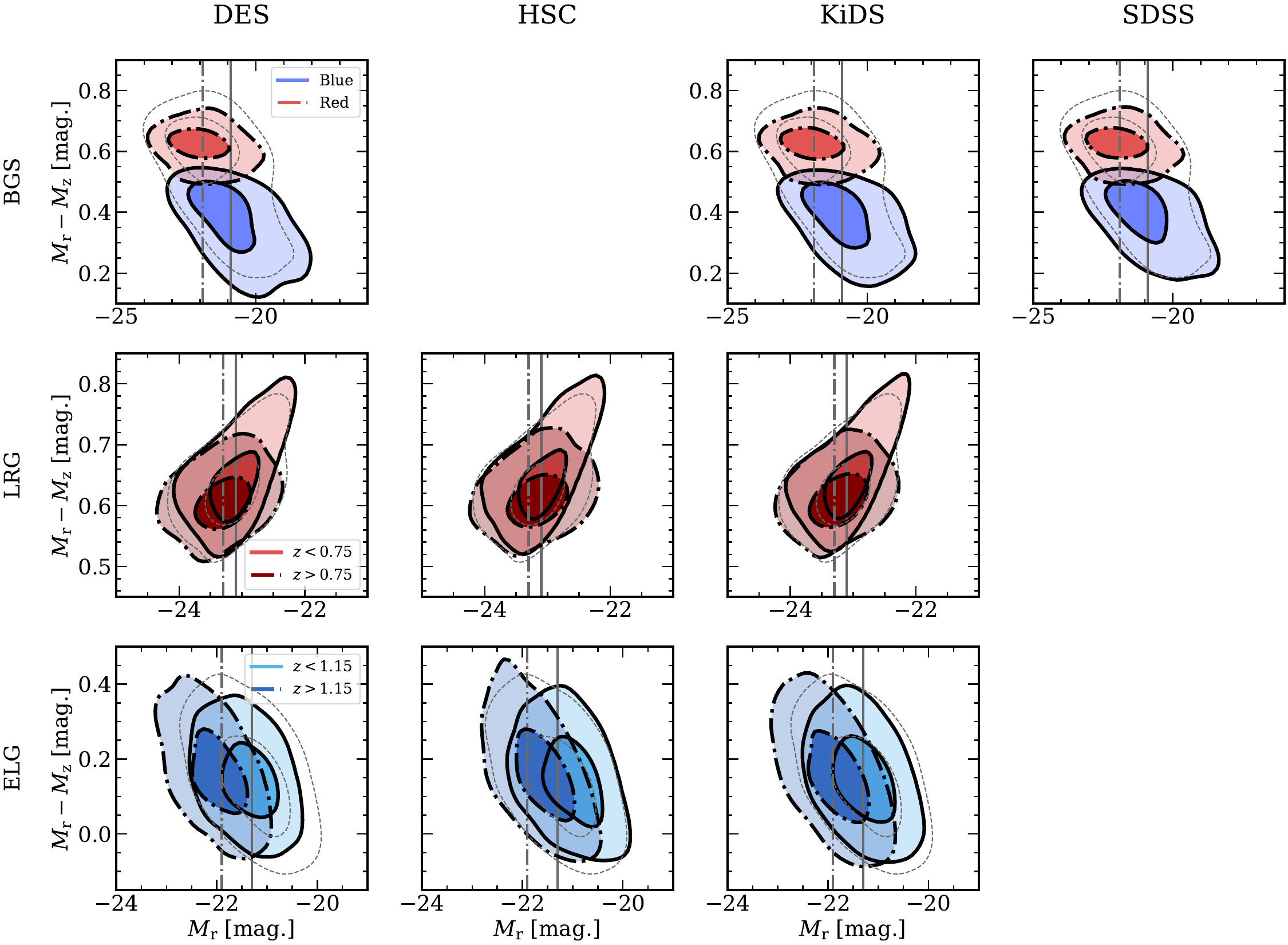}
    \caption{
    We present the absolute colours and magnitudes of our galaxy sub-samples, for each combination of imaging survey and DESI sample.
    The pre-matched DESI samples are shown in grey (dashed line).
    For BGS, the blue sample ($M_\mathrm{r}-M_\mathrm{z}<0.5$) is shown in blue with the red sample in red. 
    For the LRG and ELG samples, solid and dashed contours correspond to the low and high redshift cuts, respectively.
    The vertical lines demarcate the luminosity splits.
    The contours indicate the $39$ and $86$th percentiles.
    } 
    \label{fig:subsamples_2d}
\end{figure*}

\begin{table*}
\begin{center}
\begin{tabular}{ccc | ccccccccc}
\hline
Imaging & Sample & Population & No. of Galaxies & $f_{\rm match}$ & $\bar{n}$ & $\sigma_{\rm e}$ & $\mathcal{R}$ & $\bar{m}$ & Redshift & Colour & Luminosity  \\
 &  & &  & & [deg.$^{-2}$] & & & & $\langle z \rangle$  & $\langle M_r-M_z \rangle$  & $\langle M_r \rangle$  \\
\hline
DES & BGS & Blue + Faint & 39,081 & $0.771$ & $55.7$ & $0.2$ & $0.848$ & $-0.006$ & $0.13$ & $0.29$ & $-19.77$\\
 &  & Blue + Bright & 33,301 & $0.75$ & $47.5$ & $0.13$ & $0.825$ & $-0.006$ & $0.27$ & $0.39$ & $-21.7$\\
 &  & Red + Faint & 65,221 & $0.722$ & $93.0$ & $0.19$ & $0.734$ & $-0.006$ & $0.2$ & $0.62$ & $-21.08$\\
 &  & Red + Bright & 64,495 & $0.749$ & $92.0$ & $0.12$ & $0.693$ & $-0.006$ & $0.33$ & $0.63$ & $-22.59$\\
\cline{2-12}
  & LRG & Low $z$ + Faint & 39,597 & $0.891$ & $49.3$ & $0.18$ & $0.71$ & $-0.02$ & $0.57$ & $0.69$ & $-22.73$\\
 &  & Low $z$ + Bright & 37,974 & $0.883$ & $47.3$ & $0.11$ & $0.675$ & $-0.024$ & $0.6$ & $0.62$ & $-23.43$\\
 &  & High $z$ + Faint & 33,316 & $0.685$ & $41.5$ & $0.13$ & $0.64$ & $-0.024$ & $0.85$ & $0.62$ & $-22.99$\\
 &  & High $z$ + Bright & 33,298 & $0.859$ & $41.5$ & $0.12$ & $0.659$ & $-0.024$ & $0.9$ & $0.59$ & $-23.65$\\
\cline{2-12}
  & ELG & Low $z$ + Faint & 18,043 & $0.312$ & $24.3$ & $0.19$ & $0.589$ & $-0.037$ & $0.94$ & $0.13$ & $-20.94$\\
 &  & Low $z$ + Bright & 16,917 & $0.575$ & $22.8$ & $0.19$ & $0.7$ & $-0.037$ & $1.0$ & $0.19$ & $-21.8$\\
 &  & High $z$ + Faint & 14,239 & $0.21$ & $19.2$ & $0.2$ & $0.525$ & $-0.037$ & $1.29$ & $0.12$ & $-21.55$\\
 &  & High $z$ + Bright & 16,116 & $0.532$ & $21.7$ & $0.2$ & $0.655$ & $-0.037$ & $1.35$ & $0.22$ & $-22.42$\\
\hline
HSC & LRG & Low $z$ + Faint & 33,560 & $0.721$ & $77.6$ & $0.34$ & $0.856$ & $-0.074$ & $0.59$ & $0.69$ & $-22.73$\\
 &  & Low $z$ + Bright & 29,684 & $0.745$ & $68.6$ & $0.24$ & $0.857$ & $-0.071$ & $0.61$ & $0.62$ & $-23.42$\\
 &  & High $z$ + Faint & 39,574 & $0.746$ & $91.5$ & $0.31$ & $0.877$ & $-0.061$ & $0.86$ & $0.62$ & $-22.92$\\
 &  & High $z$ + Bright & 28,702 & $0.755$ & $66.4$ & $0.26$ & $0.865$ & $-0.068$ & $0.91$ & $0.6$ & $-23.63$\\
\cline{2-12}
  & ELG & Low $z$ + Faint & 56,283 & $0.635$ & $129.8$ & $0.39$ & $0.856$ & $-0.054$ & $0.96$ & $0.14$ & $-20.77$\\
 &  & Low $z$ + Bright & 29,484 & $0.702$ & $68.0$ & $0.36$ & $0.856$ & $-0.06$ & $1.01$ & $0.21$ & $-21.73$\\
 &  & High $z$ + Faint & 49,477 & $0.585$ & $114.1$ & $0.39$ & $0.849$ & $-0.051$ & $1.31$ & $0.14$ & $-21.36$\\
 &  & High $z$ + Bright & 27,003 & $0.674$ & $62.3$ & $0.37$ & $0.849$ & $-0.059$ & $1.36$ & $0.26$ & $-22.35$\\
\hline
KiDS & BGS & Blue + Faint & 41,112 & $0.806$ & $91.9$ & $0.33$ & --- & $0.324$ & $0.13$ & $0.32$ & $-19.85$\\
 &  & Blue + Bright & 34,257 & $0.819$ & $76.6$ & $0.17$ & --- & $0.334$ & $0.26$ & $0.4$ & $-21.68$\\
 &  & Red + Faint & 82,077 & $0.907$ & $183.5$ & $0.31$ & --- & $0.258$ & $0.18$ & $0.62$ & $-21.02$\\
 &  & Red + Bright & 71,972 & $0.847$ & $160.9$ & $0.2$ & --- & $0.278$ & $0.31$ & $0.63$ & $-22.56$\\
\cline{2-12}
  & LRG & Low $z$ + Faint & 42,827 & $0.822$ & $96.3$ & $0.26$ & --- & $0.175$ & $0.58$ & $0.69$ & $-22.74$\\
 &  & Low $z$ + Bright & 37,839 & $0.821$ & $85.0$ & $0.19$ & --- & $0.24$ & $0.61$ & $0.62$ & $-23.41$\\
 &  & High $z$ + Faint & 40,319 & $0.712$ & $90.6$ & $0.21$ & --- & $-0.075$ & $0.85$ & $0.63$ & $-22.94$\\
 &  & High $z$ + Bright & 30,464 & $0.747$ & $68.5$ & $0.2$ & --- & $0.055$ & $0.89$ & $0.6$ & $-23.64$\\
\cline{2-12}
  & ELG & Low $z$ + Faint & 39,298 & $0.388$ & $87.8$ & $0.26$ & --- & $-0.224$ & $0.94$ & $0.12$ & $-20.87$\\
 &  & Low $z$ + Bright & 34,708 & $0.704$ & $77.6$ & $0.24$ & --- & $-0.104$ & $1.0$ & $0.2$ & $-21.79$\\
 &  & High $z$ + Faint & 33,835 & $0.325$ & $75.6$ & $0.26$ & --- & $-0.231$ & $1.31$ & $0.12$ & $-21.46$\\
 &  & High $z$ + Bright & 28,853 & $0.599$ & $64.5$ & $0.25$ & --- & $-0.131$ & $1.35$ & $0.24$ & $-22.4$\\
\hline
SDSS & BGS & Blue + Faint & 254,507 & $0.674$ & $53.0$ & $0.38$ & $0.858$ & --- & $0.13$ & $0.33$ & $-19.87$\\
 &  & Blue + Bright & 257,172 & $0.837$ & $53.6$ & $0.28$ & $0.924$ & --- & $0.25$ & $0.4$ & $-21.71$\\
 &  & Red + Faint & 569,187 & $0.851$ & $118.6$ & $0.42$ & $0.82$ & --- & $0.18$ & $0.62$ & $-21.05$\\
 &  & Red + Bright & 527,096 & $0.89$ & $109.9$ & $0.31$ & $0.902$ & --- & $0.3$ & $0.63$ & $-22.58$\\
\hline
\end{tabular}
\caption{
The properties of the redshift--shape cross-matched sub-samples used in this work.
For each galaxy population we report the number of matched galaxies, the match fraction $f_{\rm match}$ (the number of galaxies in the matched sample relative to the parent DESI sample, within a $10$~deg.$^2$ area where the imaging and DESI footprints overlap), the mean surface density, the ellipticity dispersion \citep{Heymans2012}, the shear responsivity, and the weighted mean redshift, absolute colour, and absolute magnitude. 
}
\label{tab:subpopulations}
\end{center}
\end{table*}

We measure IA by shape--density correlations.
The DESI~DR1 BGS, LRG, and ELG samples constitute our density tracers, and we consider four lensing surveys for our shape measurements (DES, KiDS, HSC, and SDSS).
We require precise redshift information to measure shape--density correlations on $\lesssim 200~{\Mpch}$ scales and therefore cross-match each shear catalogue with each DESI tracer catalogue (matching tolerance of $1''$). 
The number of matches for each combination of DESI sample and lensing survey are reported in Table~\ref{tab:data_cross_matching}.
Figure~\ref{fig:samples_2d} presents the distributions of luminosity, rest-frame colour, apparent magnitude, and redshift for the DESI and DESI--lensing cross-matched samples.

The efficiency of cross-matching is particularly low for BGS--HSC and SDSS--ELG (Table~\ref{tab:data_cross_matching} and Figure~\ref{fig:samples_2d}); both combinations are omitted from our measurements. 
The HSC shear catalogue is limited to $z\gtrsim 0.3$ \citep{Li2022}, resulting in few matches with the BGS sample.
ELG--SDSS cross-matching is limited by the depth of SDSS imaging.
Similarly, the LRG--SDSS cross-matched sample is skewed towards brighter galaxies and lower redshifts;
we report the measured IA correlations in Appendix~\ref{sec:survey_to_survey_appendix} but omit the LRG--SDSS measurements from our primary analysis.

To study the dependence of the intrinsic alignment signal on galaxy colour, luminosity, and redshift, we divide the DESI--cross matched shear catalogues into sub-populations.
The cuts are outlined in Section~\ref{sec:data_samples} and visualized in Figure~\ref{fig:subsamples_2d}.
The number of galaxies in each sub-sample and their mean properties are reported in Table~\ref{tab:subpopulations}.

\section{Cross Lensing Survey Consistency and Combination}
\label{sec:survey_to_survey_appendix}

\begin{table*}
\begin{center}
\begin{tabular}{ll|ccc}
\hline
 &  & DES &  HSC & KiDS\\
\hline
BGS & KiDS & $3.9\times10^{-1}$ $(0.9\sigma$) & --- & ---\\
 & SDSS & $9.3\times10^{-1}$ $(0.1\sigma$) & --- & $3.9\times10^{-1}$ $(0.9\sigma$)\\
\hline
LRG & HSC & $9.5\times10^{-1}$ $(0.1\sigma$) & --- & ---\\
 & KiDS & $7.7\times10^{-1}$ $(0.3\sigma$) & $6.2\times10^{-1}$ $(0.5\sigma$) & ---\\
\hline
ELG & HSC & $6.2\times10^{-3}$ $(2.7\sigma$) & --- & ---\\
 & KiDS & $7.8\times10^{-2}$ $(1.8\sigma$) & $4.3\times10^{-1}$ $(0.8\sigma$) & ---\\
\hline
\end{tabular}
\caption{
For each DESI tracer, we report the p-values and number of standard deviations from the null hypothesis for a one-sided Gaussian from a $\chi^2$ test between each lensing survey's {\wgp} signal (between $6$ and $65~{\Mpch}$). 
}
\label{tab:survey_to_survey}
\end{center}
\end{table*}

\begin{figure*}
    \includegraphics[width=\textwidth]{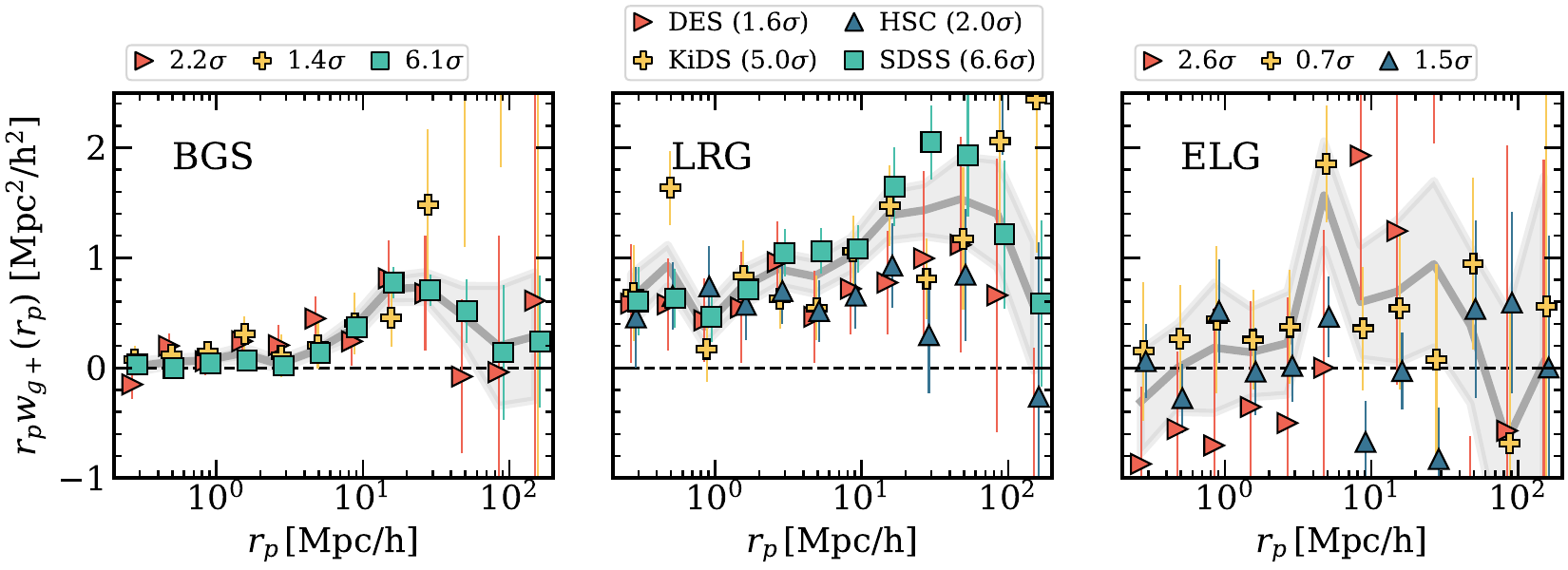}
    \caption{ 
    Measured intrinsic alignment correlations for three galaxy samples: DESI~DR1 BGS, LRG, and ELG. 
    Each colour corresponds to a different survey as the shape tracer (e.g., DES, KiDS, HSC, and SDSS).
    The significance of a $\chi^2$ test (with zero signal as the null hypothesis) is reported for each measurement, in terms of standard deviations for a one-sided Gaussian.
    For a given galaxy sample, the signal co-added across the imaging surveys is presented as a solid grey line, with the $1\sigma$ uncertainties shown as a shaded region; HSC is omitted from co-adding due to its significant overlap with the other imaging surveys.
    } 
    \label{fig:measurement_summary_wgp_overplotted}
\end{figure*}

We find that the intrinsic alignment measurements are statistically consistent between the imaging surveys. This is consistent with the findings of \citet{heydenreich2025}, which investigates the consistency of galaxy-galaxy lensing signals from the same data. However, given the dependence of direct IA measurements on the shape measurement methodology from different imaging surveys, the consistency across galaxy-galaxy lensing measurements did not necessarily require the same of IA.
Figure~\ref{fig:measurement_summary_wgp_overplotted} presents the measured IA signals for the three DESI samples from all four lensing surveys. 
For a given DESI tracer, we calculate the $\chi^2$ between each lensing survey's $w_{\rm g+}$ measurement and those of the other surveys;
$\chi^2$ is calculated relative to the null hypothesis of zero signal between $6-65~{\Mpch}$ (corresponding to $4$ degrees of freedom).
The $\chi^2$ test p-values are reported in Table~\ref{tab:survey_to_survey};
the p-values are also presented in terms of standard deviations for a one-sided Gaussian.

Given the consistency between the surveys, throughout this paper we present the co-added IA measurement across the shear catalogues. 
Co-adding is performed as
\begin{align}
    w_{\rm g+}(r_{\rm p}) = \frac{\sum_i w_{g+,i}(r_{\rm p}) / \sigma_{ w_{g+,i}(r_{\rm p}) }^2}{ \sum_i 1 / \sigma_{ w_{g+,i}(r_{\rm p}) }^2 },\\
    \sigma_{ w_{g+,i}(r_{\rm p}) } = \left [ \sum_i  1/ \sigma_{ w_{g+,i}(r_{\rm p}) }^2 \right ]^{-1},
\end{align}
where $ w_{g+,i}(r_{\rm p})$ is the measured shape--density correlation for the $i$th shear catalogue and $\sigma_{ w_{g+,i}(r_{\rm p}) }$ is the corresponding uncertainty.
For the NLA and TATT modelling, we do not consider the co-added IA correlations but instead jointly model the individual measurements.

\section{Two-point Clustering}
\label{sec:clustering_appendix}
\begin{figure}
 \includegraphics[width=0.87\columnwidth]{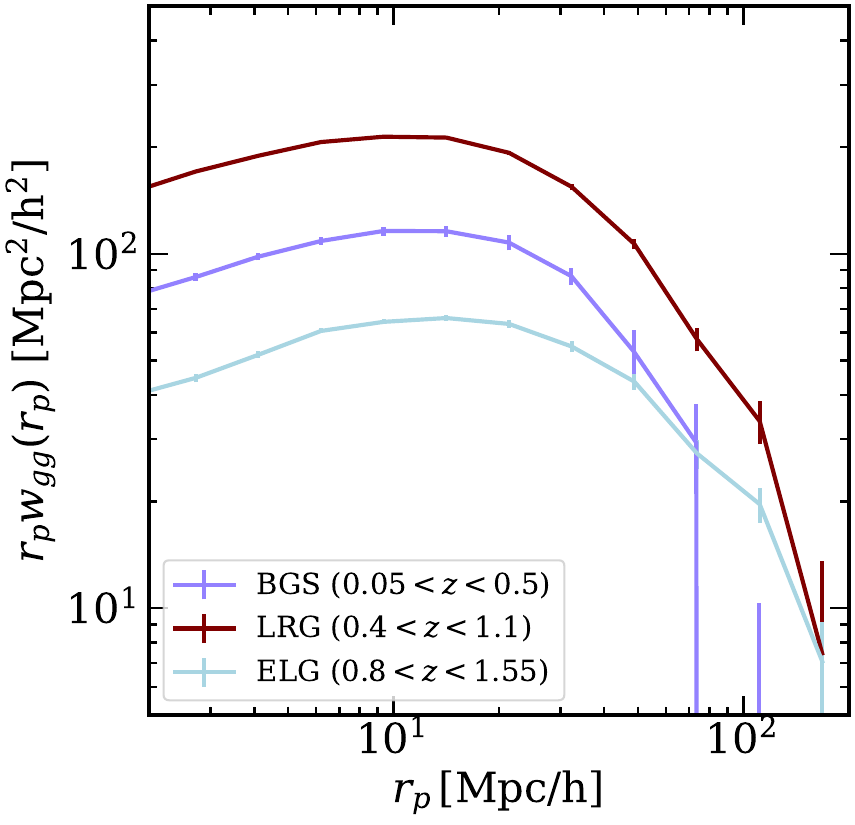}
    \caption{ 
    Measured galaxy clustering for the three DESI density tracers across the full DESI DR1 footprint: BGS, LRG, and ELG.
    } 
    \label{fig:wgg}
\end{figure}

We report the two point clustering of the DESI main survey tracer populations in Fig.~\ref{fig:wgg}.
Modelling of the clustering signal is discussed in Jeffrey et al. (in prep).

\section{Measurement Covariances}
\label{sec:covariances_appendix}

We estimate the covariance matrix of each two-point measurement via a three-dimensional leave-one-out jackknife process \citep[e.g.,][]{Johnston_2019, Samuroff_2023, HervasPeters2024}.
For a given imaging survey and DESI sample, we divide the on-sky footprint into two-dimensional patches and subdivide each on-sky region into three-dimensional slices. 
We then leave out one jackknife slice at a time and measure the two-point correlation function on the remaining catalogue.
The covariance matrix is estimated from the collection of leave-one-out measurements $\{ \mathbf{w}_{1, ab}, \dots, \mathbf{w}_{N, ab}  \} $ as
\begin{equation}
    \mathbf{C} = \frac{N-1}{N} \sum_{i=0}^N ( \mathbf{w}_{i, ab} - \bar{\mathbf{w}}_{\rm ab} ) ( \mathbf{w}_{i, ab} - \bar{\mathbf{w}}_{\rm ab} )^T,
\end{equation}
where $N$ is the number of three-dimensional jackknife regions.

\begin{figure}
\includegraphics[width=0.9\columnwidth]{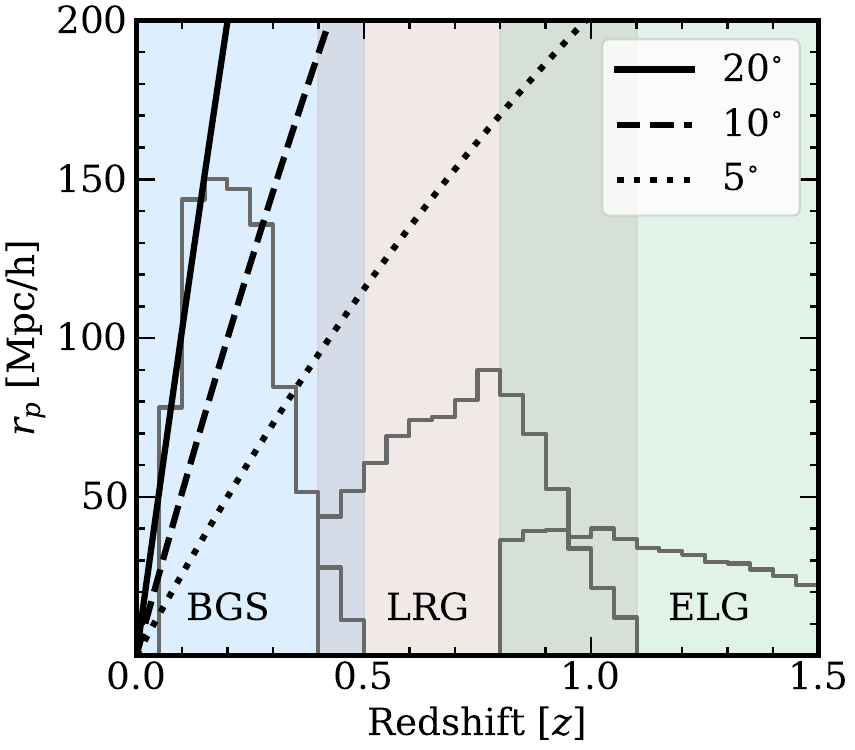}
\caption{
The relationship between transverse comoving distance and angular scale, as a function of redshift. 
The weighted redshift distributions of the KiDS--DESI cross-matched shape catalogues are included for reference; KiDS is shown as a representative example.
To measure the two-point correlation function to transverse comoving distances of $200~{\Mpch}$, the jackknife regions must be at least $20, 10,$~and$~5$~degrees for BGS, LRG, and ELG samples, respectively.
} 
\label{fig:jk}
\end{figure}

The size of the three-dimensional jackknife regions must be greater than the largest physical scale of interest in the $\mathbf{w}_{\rm ab}$ measurement; otherwise, the variance in the signal at the largest scales will be underestimated \citep[see Appendix A of][]{Johnston_2019}.   
Since we calculate the projected correlation function by integrating the three-dimensional correlation function along the line-of-sight between $-\Pi_\mathrm{max}$ and $\Pi_\mathrm{max}$ (see Equation~\ref{eqn:twopoint}), the jackknife regions must be greater than $2 \Pi_\mathrm{max}$ in depth. 
The minimum size of jackknife regions on the sky depends on the redshift distribution of the sample.
Figure~\ref{fig:jk} presents the mapping between angular scale on the sky and transverse comoving distance as a function of redshift.
Measuring $\mathbf{w}_{\rm ab}$ to transverse comoving distances of $200~{\Mpch}$ requires jackknife regions of at least $20, 10,$~and$~5$~degrees for BGS, LRG, and ELG samples, respectively.

\section{Scale dependence of the fits}
\label{sec:scale_dep_appendix}

In this paper, we model IA measurements with the NLA and TATT models.
NLA neglects higher-order correlations and is only valid on large scales ($\gtrsim6$~{\Mpch}), while TATT is typically applied to smaller scales ($\gtrsim2$~{\Mpch}).
To test the ability of the TATT model to fit the data on small and large scales, in Table~\ref{tab:NLA_TATT_ext} we report parameter posteriors and goodness of fit for TATT fits to $6$ galaxy samples using a variety of scale cuts;
we also include the fiducial NLA fits for reference.
The galaxy bias estimates are in general agreement with previous DESI studies (e.g., \citealt{desi_edr_clustering}).
The models are fit to four selections of BGS galaxies (the same samples as Figure~\ref{fig:TATT}): three blue galaxy samples defined in terms of restframe colour $ M_\mathrm{r} - M_\mathrm{z}$ cuts and one sample of red galaxies for comparison. 
DES, KiDS, and SDSS are the shape tracers for the BGS measurements.
Low-$z$ ($0.8<z<1.15$) and high-$z$ ($1.15<z<1.55$) ELG samples are also presented;
HSC is the shape tracer.
The TATT fits are largely independent of the applied scale cut.  
A full investigation of scale cuts for the NLA and TATT models will be provided by Jeffrey et al. in prep.

\begin{table*}
\begin{center}
\begin{tabular}{cccclllllcc}
\hline
Sample & Selection & Model & Min. scale & $A_1 (A_\mathrm{IA})$ & $A_2$ & $b_\mathrm{TA}$ & $b_1$ & $b_2$ & {\wgg} $\chi^2_\nu$ & {\wgp} $\chi^2_\nu$\\
\hline
BGS & $M_r-M_z<0.5$ & NLA & 6 &  $0.1_{-0.3}^{+0.3}$ & ---  & ---  & $1.142_{-0.004}^{+0.004}$ & ---  & $0.83$ & $1.0$ \\
\cline{3-11}
 &  & TATT & 6 &  $0.7_{-0.6}^{+0.7}$ & $-1.0_{-2.0}^{+1.0}$ & $0.0_{-1.0}^{+1.0}$ & $1.1_{-0.02}^{+0.02}$ & $-0.7_{-0.2}^{+0.2}$ & $0.65$ & $1.29$ \\
 &  &  & 2 &  $0.2_{-0.4}^{+0.4}$ & $-0.1_{-0.3}^{+0.6}$ & $-1.0_{-1.0}^{+1.0}$ & $1.145_{-0.003}^{+0.003}$ & $-0.05_{-0.03}^{+0.03}$ & $0.78$ & $2.13$ \\
\cline{2-11}
 & $M_r-M_z<0.6$ & NLA & 6 &  $0.9_{-0.2}^{+0.2}$ & ---  & ---  & $1.149_{-0.004}^{+0.004}$ & ---  & $0.87$ & $0.93$ \\
\cline{3-11}
 &  & TATT & 6 &  $0.9_{-0.3}^{+0.4}$ & $0.0_{-2.0}^{+1.0}$ & $-0.0_{-1.0}^{+1.0}$ & $1.11_{-0.02}^{+0.02}$ & $-0.7_{-0.2}^{+0.3}$ & $0.69$ & $1.32$ \\
 &  &  & 2 &  $0.8_{-0.2}^{+0.3}$ & $0.8_{-0.5}^{+0.4}$ & $-1.5_{-0.3}^{+0.4}$ & $1.151_{-0.003}^{+0.003}$ & $-0.04_{-0.04}^{+0.03}$ & $0.8$ & $1.11$ \\
\cline{2-11}
 & $M_r-M_z<0.65$ & NLA & 6 &  $1.4_{-0.2}^{+0.2}$ & ---  & ---  & $1.153_{-0.004}^{+0.004}$ & ---  & $0.89$ & $2.12$ \\
\cline{3-11}
 &  & TATT & 6 &  $1.8_{-0.3}^{+0.5}$ & $1.0_{-2.0}^{+1.0}$ & $-1.0_{-0.7}^{+0.9}$ & $1.11_{-0.02}^{+0.02}$ & $-0.7_{-0.2}^{+0.3}$ & $0.71$ & $1.42$ \\
 &  &  & 2 &  $1.4_{-0.3}^{+0.3}$ & $1.1_{-0.5}^{+0.5}$ & $-1.2_{-0.3}^{+0.2}$ & $1.156_{-0.004}^{+0.003}$ & $-0.04_{-0.04}^{+0.03}$ & $0.84$ & $1.9$ \\
\hline
ELG & Low-$z$ & NLA & 6 &  $-0.0_{-1.0}^{+1.0}$ & ---  & ---  & $1.254_{-0.006}^{+0.006}$ & ---  & $0.64$ & $0.74$ \\
\cline{3-11}
 &  & TATT & 6 &  $-0.0_{-2.0}^{+2.0}$ & $0.0_{-5.0}^{+5.0}$ & $-0.0_{-1.0}^{+1.0}$ & $1.254_{-0.01}^{+0.007}$ & $0.1_{-0.5}^{+0.4}$ & $0.99$ & --- \\
 &  &  & 2 &  $-1.0_{-1.0}^{+2.0}$ & $3.0_{-4.0}^{+4.0}$ & $-0.0_{-1.0}^{+1.0}$ & $1.25_{-0.005}^{+0.006}$ & $-0.2_{-0.2}^{+0.2}$ & $0.75$ & $3.04$ \\
\cline{2-11}
 & High-$z$ & NLA & 6 &  $-3.0_{-1.0}^{+1.0}$ & ---  & ---  & $1.452_{-0.007}^{+0.007}$ & ---  & $1.57$ & $1.18$ \\
\cline{3-11}
 &  & TATT & 6 &  $-1.0_{-2.0}^{+2.0}$ & $-5.0_{-4.0}^{+6.0}$ & $0.0_{-1.0}^{+1.0}$ & $1.45_{-0.009}^{+0.007}$ & $0.1_{-0.5}^{+0.5}$ & $2.27$ & --- \\
 &  &  & 2 &  $-2.0_{-3.0}^{+2.0}$ & $-1.0_{-5.0}^{+5.0}$ & $-1.0_{-1.0}^{+2.0}$ & $1.443_{-0.006}^{+0.006}$ & $-0.3_{-0.3}^{+0.2}$ & $1.57$ & $4.22$ \\
\hline
\end{tabular}
\caption{
Parameter posteriors and reduced $\chi^2$ are reported for NLA and TATT fits.
The IA measurements are identical to Table~\ref{tab:NLA_TATT}.
For the BGS measurements, DES, KiDS, and SDSS are jointly fit.
For the ELG measurements, HSC is the only shape tracer.
The NLA model is fit between $6$ and $65$~{\Mpch} and the TATT model is fit between either $2-65$~{\Mpch} or $6-65$~{\Mpch}.
Reduced $\chi^2$ is undefined for the $>6$~{\Mpch} ELG TATT fits, because the number of {\wgp} data-points is smaller  than the number of free parameters.
}
\label{tab:NLA_TATT_ext}
\end{center}
\end{table*}

\section{On the purity of blue galaxy samples---dependence on luminosity}
\label{sec:blue_luminosity}

\begin{figure*}
    \includegraphics[width=\textwidth]{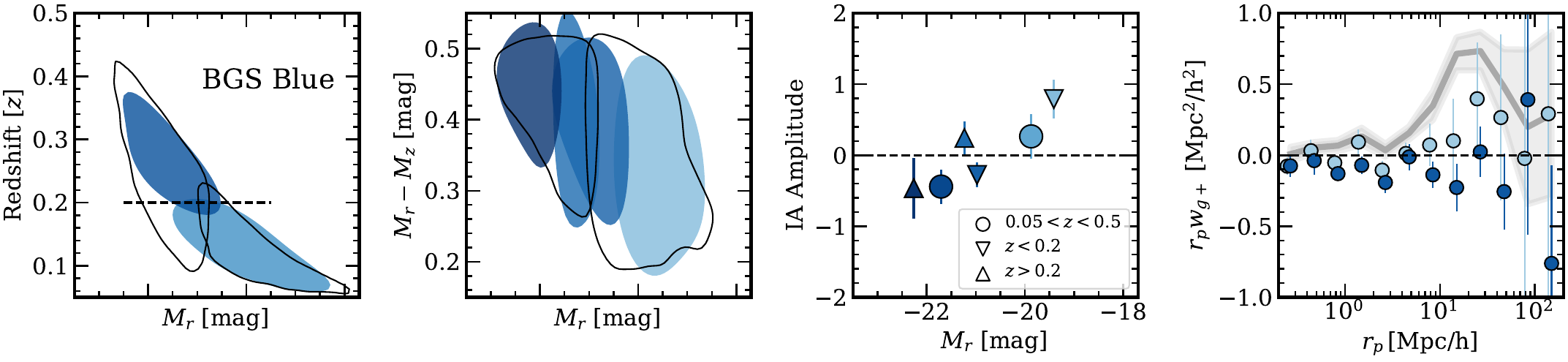}
    \caption{
    The NLA $A_\mathrm{IA}$ amplitude of  blue BGS galaxies in luminosity and redshift bins.
    For blue ($M_\mathrm{r} - M_\mathrm{z} < 0.5$) BGS galaxies, we divide the SDSS shear catalogue into four sub-samples: first, a low and high redshift bin at $z=0.2$; the first panel shows the distributions of redshift and luminosity for each bin. 
    Next, each redshift bin is divided into two luminosity bins; the second panel shows the distributions of rest frame colour and luminosity for each bin.  
    The colour scale corresponds to rest-frame colour.
    The third panel presents the NLA amplitude of the intrinsic alignments as a function of mean luminosity, for each bin.
    For reference, we also show the blue-faint and blue-bright samples defined in Section~\ref{sec:data_samples} and shown in Figure~\ref{fig:subsample_summary}, using the full BGS redshift range and wider luminosity bins; these samples are shown as black contours on the first two panels and as circles in the third panel.
    The rightmost panel presents the IA signal for the blue-faint and blue-bright samples, alongside the parent BGS measurements (i.e., no colour, luminosity, or redshift cuts).
    Note that DES, KiDS, and SDSS are the shape tracers for the blue-faint and bright samples; only SDSS is the shape tracer for the four narrower bins.
    } 
    \label{fig:blue_amplitude}
\end{figure*}

In Section~\ref{sec:blue_purity}, we investigated how the IA signal depends on the severity of blue galaxy selection.
Here we investigate whether additional selection criteria are warranted. 

We divide the BGS-matched \textit{blue} ($M_\mathrm{r} - M_\mathrm{z} < 0.5$) shape catalogues into four independent bins:
we define two redshift ranges---$0.05<z<0.2$ and $0.2<z<0.5$---and divide each into two luminosity bins.
Due to the narrowness of the bins, 
SDSS is the only shape tracer.
The distribution of luminosity, rest-frame colour, and redshift for each bin is presented in Figure~\ref{fig:blue_amplitude}.
We also consider the blue-faint and blue-bright samples defined in Section~\ref{sec:data_samples};
these larger samples do not include a redshift cut and have DES, KiDS, and SDSS shapes.
As a magnitude limited sample, luminosity and redshift are strongly correlated for the BGS galaxies;
the fainter samples are skewed towards lower redshifts.

For each BGS-matched blue sample, we measure the IA signal relative to the BGS density tracers.
Figure~\ref{fig:blue_amplitude} presents the NLA amplitude, {\AIA}, for each bin, as well as the measured $w_{\rm g+}(r_{\rm p})$ signal for the blue-faint and blue-bright samples.
All measurements are consistent with zero at $3\sigma$, however, there is an apparent trend between amplitude and luminosity: the faintest blue galaxies appear more aligned than the brightest blue galaxies.
The measured signals are low amplitude ($A_{\rm IA}\lesssim1$) and statistically low significance.
Due to the strong correlations between colour, luminosity, and redshift for our samples, the providence of this trend is unclear.
If luminosity is the driving factor, we are potentially observing the alignment of faint blue satellites with their local environment. 
However, for these narrow bins in luminosity and redshift, SDSS is our only shape tracer, and unknown imaging systematics may be skewing our results.
Future studies of blue faint galaxies with LSST will be critical to confirming these potential trends. 

\label{lastpage}												


\end{document}